\documentclass[fleqn,usenatbib]{mnras}

\usepackage{newtxtext,newtxmath}
\usepackage[T1]{fontenc}

\DeclareRobustCommand{\VAN}[3]{#2}
\let\VANthebibliography\thebibliography
\def\thebibliography{\DeclareRobustCommand{\VAN}[3]{##3}\VANthebibliography}

\usepackage{graphicx}	% Including figure files
\usepackage{amsmath}	% Advanced maths commands
\title[Extragalactic radio sources behind the MCs]{Identification of extragalactic radio sources in the VISTA survey of the Magellanic Clouds using an unsupervised machine learning algorithm\thanks{based on observations made with the Southern African Large Telescope}}
\author[C. M. Pennock et al.]{
Clara M. Pennock,$^{1}$\thanks{E-mail: Clara.Pennock@ed.ac.uk}
Jacco Th. van Loon,$^{2}$
Mohammed Iddrisu Nlowie,$^{1}$
Elizabeth Joy Leese,$^{3}$
\newauthor Maja Nowaczyk,$^{4}$
Heinz Andernach,$^{5,6}$
Duncan Farrah,$^{7,8}$
Valentin D. Ivanov,$^{9}$
Micha{\l}~J. Micha{\l}owski,$^{10}$
\newauthor Joana M. Oliveira,$^{2}$
Jahang Prathap,$^{11,12,13}$
Syed F. Rahman,$^{14,15}$
Mara Salvato,$^{16,17}$
Tayyaba Zafar$^{11,12}$
\\
$^{1}$Institute for Astronomy, University of Edinburgh, Royal Observatory, Blackford Hill, Edinburgh EH9 3HJ, UK \\
$^{2}$School of Engineering and Physical Sciences, Lennard-Jones Laboratories, Keele University, ST5 5BG, UK \\
$^{3}$St. Margaret Ward Catholic Academy, Little Chell Lane, Tunstall, Stoke-on-Trent, ST6 6LZ, UK \\
$^{4}$St. John Fisher Catholic College, Ashfields New Rd, Newcastle-under-Lyme, ST5 2SJ, UK \\
$^{5}$Depto.\ de Astronom\'{i}a, DCNE, Universidad de Guanajuato, Callej\'on de Jalisco s/n, Guanajuato, C.P. 36023, Mexico \\
$^{6}$Th\"uringer Landessternwarte, Sternwarte 5, D-07778 Tautenburg, Germany \\
$^{7}$Department of Physics and Astronomy, University of Hawai‘i at Mānoa, 2505 Correa Rd., Honolulu, HI96822,USA \\
$^{8}$Institute for Astronomy, University of Hawai‘i, 2680 Woodlawn Dr., Honolulu, HI96822, USA \\
$^{9}$ European Southern Observatory, Karl-Schwarzschild-Strasse 2, D-85748 Garching bei M\"unchen, Germany\\
$^{10}$Astronomical Observatory Institute, Faculty of Physics and Astronomy, Adam Mickiewicz University, ul. S{\l}oneczna 36, 60-286, Pozna\'{n}, Poland \\
$^{11}$School of Mathematical and Physical Sciences, Macquarie University, Sydney, NSW 2109, Australia\\
$^{12}$Astrophysics and Space Technologies Research Centre, Macquarie University , Sydney, NSW 2109, Australia\\
$^{13}$Australia Telescope National Facility, CSIRO, Space and Astronomy, PO Box 1130, Bentley, WA 6102, Australia\\
$^{14}$Syed Babar Ali School of Science and Engineering, Lahore University of Management Sciences, Lahore, Pakistan \\
$^{15}$NCBC at NED University of Engineering and Technology, Karachi, Pakistan\\
$^{16}$Max-Planck-Institut für extraterrestrische Physik, Giessenbachstr. 1,
85748 Garching, Germany\\
$^{17}$ Exzellenzcluster ORIGINS, Boltzmannstr. 2, D-85748 Garching,
Germany\\
}

\date{Accepted XXX. Received YYY; in original form ZZZ}

\pubyear{2025}

\begin{document}
\label{firstpage}
\pagerange{\pageref{firstpage}--\pageref{lastpage}}
\maketitle

% Abstract of the paper
\begin{abstract}
%This is a simple template for authors to write new MNRAS papers.
%The abstract should briefly describe the aims, methods, and main results of the paper.
%It should be a single paragraph not more than 250 words (200 words for Letters).
The Magellanic Clouds benefit from a wealth of data across the electromagnetic spectrum, but extracting extragalactic sources from these crowded fields is a less than straightforward endeavour. 
We used an unsupervised machine learning algorithm (t-distributed stochastic neighbourhood embedding; t-SNE) to classify 6,652 and 1,356 sources in the Large and Small Magellanic Clouds, respectively, that were detected in the VISTA Survey of the Magellanic Clouds (VMC) and the Evolutionary Map of the Universe (EMU) survey with the Australian Square Kilometre Array Pathfinder (ASKAP). Multi-wavelength photometry and colours from optical to radio were used as features to cluster similar sources together in 2D space on a t-SNE map. We used previously classified sources to separate the clusters into five classes: AGN, low-redshift galaxies, blazars, dusty stellar objects (e.g. young stellar objects, planetary nebul\ae\, etc) and high proper-motion stars. 
Previous photometric classifications from the literature match with specific clusters identified, as well as provide insight on areas of the t-SNE map for which there are no previous classifications. 
This work shows that the combination of unsupervised machine learning with photometry from optical to radio has proven adept at separating the stellar from the extragalactic, and the AGN from the galaxies. Our associated new spectroscopy with SALT includes new discoveries of AGN up to redshift 3.3 and Magellanic emission-line objects.
\end{abstract}

% Select between one and six entries from the list of approved keywords.
% Don't make up new ones.

%%%%%%%%%%%%%%%%%%%%%%%%%%%%%%%%%%%%%%%%%%%%%%%%%%

%%%%%%%%%%%%%%%%% BODY OF PAPER %%%%%%%%%%%%%%%%%%

\section{Introduction}
Identifying extragalactic sources from sky fields crowded with stars, such as the Galactic Plane and the Magellanic Clouds, is a less than straightforward and time consuming endeavour. As wider and deeper photometric surveys become available in these parts of the sky allowing us to better sample source's spectral energy distribution (SED), this challenge becomes viable. However, various factors, such as emission lines, dust and variability of a source's flux between measurements at different wavelengths, can cause some stellar sources to masquerade as extragalactic sources and vice versa \citep[e.g.][]{Hony2011, 2022MNRAS.515.6046P}. 

Extragalactic sources, in particular active galactic nuclei (AGN), are capable of emitting across the entire electromagnetic spectrum. Emission observed at different wavelengths relate to different parts of an AGN's structure, such as the accretion disk observed in the optical/UV, the dusty torus (or obscuring structure) in the infrared and the presence jets and/or lobes in the radio. Photometry in the infrared is a powerful tool in identifying AGN in particular as the dusty torus has a distinct bump-like shape in an SED which helps isolate these sources in colour--colour and colour--magnitude diagrams \citep[e.g.][]{Spitzer, Stern2012, 2012Mateos, Assef2013, Nikutta2014}. A radio detection is also a strong sign that we are observing an AGN, though it should be noted that star formation in galaxies and stellar sources (e.g. planetary nebul\ae\, young stellar objects) are also capable of emitting in the radio, though they tend to be much weaker than AGN and exhibit flatter radio spectral slopes \citep[e.g.][]{Joseph2019}.  

The Magellanic Clouds (including the Magellanic Bridge that links them) span $\sim$ 100 sq.\ degrees on the sky and benefit from a plethora of all-sky and Magellanic specific surveys from X-ray to radio. They have been observed in the near-infrared (near-IR) with the VISTA Magellanic Clouds survey \citep[VMC;][]{2011A&A...527A.116C} which showcases a great improvement in depth and angular resolution compared to previous IR surveys, and has detected stars encompassing most phases of evolution such as main sequence stars, red giant branch
(RGB) stars, red clump stars, RR Lyr\ae\ and Cepheid variables, asymptotic giant branch (AGB) stars, post-AGB stars, planetary nebul\ae\ (PNe) and supernova remnants (SNRs) \citep[e.g.][]{2012AGB,2013Cioni,2015Cepheid,2018YS,2019RGB,2020YS,2020AGB,2021RRL,2021RGB}.
The VMC has also had success in discovering background extragalactic sources \citep{2013Cioni, Ivanov2016, 2022MNRAS.515.6046P, Ivanov2024}. This is due to the benefits from all-sky surveys such as WISE \citep{WISE, Cutri2013} and Magellanic Clouds specific surveys with the Spitzer Space Telescope as part of the Spitzer Agents of Galaxy Evolution (SAGE) survey of the LMC \citep{meixner2006} and SMC \citep{gordon2011}, both at mid-IR wavelengths that have been used to identify AGN \citep[e.g.][]{2004Lacy,2012Donley, Stern2012}. Furthermore, the Clouds have been observed in the radio as part of the Evolutionary Map of the Universe all-sky \citep[EMU;][]{Joseph2019,2021MNRAS.506.3540P} survey. The combination of all these data has great potential for discovery of the more unusual and extreme cases of AGN, such as SAGE0536AGN \citep{vanloon2015,2022MNRAS.515.6046P}, which is a dusty AGN with one of the strongest 10$\mu$m silicate emissions currently known.

The area of the Magellanic Clouds, contains hundreds of millions of sources to sort through, which would be an extremely time-consuming task. Machine learning has been put to good use in astronomy to learn patterns in large datasets and separate different classes of sources \citep[e.g.][]{1992MNRAS.259P...8S, 2009Farrah,2012MNRAS.419.2633G, 2015MNRAS.450.1441D,2020MNRAS.491.1408M, 2022MNRAS.515.6046P, 2023A&A...674A..14R, Pennock2025, 2026PASA...43...47P}. Supervised machine learning, such as applied to the Magellanic Clouds in \citet{Pennock2025}, uses a training set from which to learn patterns which can then be applied to unseen data. However, such techniques only classify sources into previously known categories, and does not deal well with the fluctuations in real astrophysical numbers for the different classes. Unsupervised machine learning requires no training set and learns from all the data to structure and separate sources. Unsupervised machine learning is used to be able to combine a multitude of features that are not necessarily easily combined through human comparison between different diagrams and selection functions, which means it can identify groupings that are unbiased from human selection upfront. In \cite{2022MNRAS.515.6046P} an unsupervised machine learning algorithm called t-SNE (t-distributed stochastic neighbourhood embedding) was used to cluster similar sources to find sources behind the SMC that were similar to the unusual AGN, SAGE0536AGN and SAGE0534AGN, whose emission appeared to be mostly from their AGN. This paper is a continuation of the t-SNE work but will look at all the sources, instead of a small subset, as well as expanding this technique to the LMC, in the aim of exploring the populations of radio detected sources, especially the AGN, in the crowded field of the Magellanic Clouds that have a large variety of different object types, from nebulae to high-z AGN, with a particular focus on what the near-IR and radio bands can do to separate these different populations.

The structure of this paper is as follows: In Section \ref{Sec:Data} we describe the photometry catalogues available in the field of the Magellanic Clouds and how the multi-wavelength dataset is created. We describe the machine learning used, the t-SNE algorithm in Section \ref{sec:tSNE}. In Section \ref{sec:results} we present the 2D maps produced by the t-SNE algorithm and how different classifications and features of the data vary across the maps. Next, in Section \ref{sec:disc} we explore how features, such as X-ray flux and redshift, that were not used in the machine learning vary across the maps (Section \ref{extrafeats}), as well as looking at how the quality of the AllWISE photometry has affected the machine learning (Section \ref{sec:AllWISEflags}) and how this method compares to previous classification methods (Section \ref{sec:otherclass}). We also discuss follow up spectroscopy of sources found in the blazar candidate sequences (Section \ref{blazseq}) of the t-SNE maps and how the classifications of these sources match up to features used and not used in the tSNE algorithms and to classifications from previous methods. Lastly, in Section \ref{sec:conc} we present the conclusions.

Unless stated otherwise all quoted photometric magnitudes use the Vega photometric system.
%%%%%%%%%%%%%%%%%%%%%%%%%%%%%%%%%%%%%%%%%%%%%%%%%%%%%%%%%%%%%%%%%%%%%%

%%%%%%%%%%%%%%%%%%%%%%%%%%%%%%%%%%%%%%%%%%%%%%%%%%%%%%%%%%%%%%%%%%%%%%

\section{Data}\label{Sec:Data}
For both supervised and unsupervised machine learning "features" for each object, such as photometric measurements in various wavebands, are required. In this section the photometric surveys from which the features are obtained are outlined. 

\subsection{Photometry}
We aimed to choose surveys that provided a large coverage of the VISTA Magellanic Clouds and the ASKAP SMC and LMC footprints, as we are primarily interested in the synergy of near-IR and radio for detecting the extragalactic in stellar dense fields, while still reaching reasonable depths. This removed some surveys that only covered portions of the Clouds such as Spitzer SAGE \citep{meixner2006, gordon2011, SAGEdat} and XMM-Newton \citep{2013XMM} X-ray imaging surveys, as well as shallower surveys such as the Herschel HERITAGE \citep{meixner2010} survey.

\subsubsection{Near-IR VISTA Magellanic Clouds survey}

The Visible and Infrared Survey Telescope for Astronomy \citep[VISTA;][]{2015A&A...575A..25S} telescope is a 4.1-m near-infrared optimised telescope, on which was mounted the VISTA InfraRed CAMera \citep[VIRCAM;][]{2006vista}, equipped with a large array of 16 infrared detectors that cover a field of about a 1.5 square degree.

The VISTA Survey of the Magellanic Clouds \citep[VMC;][]{2011A&A...527A.116C} is a near-IR deep, multi-epoch and wide-field imaging survey of the Magellanic Clouds, covering an area of about 184 deg$^{2}$. VISTA observations for the VMC main survey started in November 2009 and ended in October 2018, with a further 12 additional programmes that provided multi-epoch data up to early 2023 \citep{2025VMCDR7}. It has a spatial resolution of 1$\rlap{.}^{\prime\prime}$0 -- 1$\rlap{.}^{\prime\prime}$1, 0$\rlap{.}^{\prime\prime}$9 -- 1$\rlap{.}^{\prime\prime}$0 and 0$\rlap{.}^{\prime\prime}$8 -- 0$\rlap{.}^{\prime\prime}$9 in the $YJK$\textsubscript{s} filters, respectively, where the two values specified for seeing are for uncrowded and crowded regions, respectively. It reaches a sensitivity of about 22, 22 and 21.5 mag (Vega) in the $YJK$\textsubscript{s} bands, respectively. 
The VMC have already proven successful in discovering more AGN \citep[e.g.][]{Ivanov2016, Ivanov2024}. This however comes with the caveat of increased stellar confusion with the presence of the LMC and SMC.

The catalogues created from the VMC survey provide both aperture and PSF photometry, where PSF photometry reaches sources on average 0.3 magnitudes fainter than aperture photometry. The PSF catalogue is created as described in \cite{2015Rubele}. The magnitudes in each band are calculated from deep tile images, which are a combination of single exposure images from different epochs. We use the PSF photometry because of its increased depth and ability to distinguish sources in crowded regions, which contain $\sim$28.6 million and $\sim$103.2 million sources in the SMC and LMC, respectively. Also provided are sharpness values in each of the bands, which is a measure of the difference between the observed width of the object and the width of the PSF model, where stars should have a sharpness value of 0 and resolved objects sharpness values of $>$0. Sharpness values $<$ 0 indicate artefacts such as bad pixels or cosmic ray impacts.

\subsubsection{Radio ASKAP survey}

To probe the extragalactic population we restrict our sample to those that are radio detected, as most of the radio population is expected to be extragalactic \citep[e.g.][]{2021MNRAS.506.3540P}. We expect that a large fraction of the near-IR galaxy population will remain undetected in radio, but many of the AGN will be detected and stellar contamination is minimised.

The Evolutionary Map of the Universe \citep[EMU;][]{norris2011, 2025EMU} is a wide-field radio continuum survey which uses the Australian Square Kilometre Array Pathfinder \citep[ASKAP;][]{Johnston2008,Hotan2021} telescope. EMU's primary goal is to make a deep (rms $\sim$ 30 $\mu$Jy/beam) radio continuum survey of the Southern sky, extending as far North as +10$^{\circ}$  declination, with a resolution of 13$^{\prime\prime}$. It is expected to catalogue about 35 million radio sources (most of which will most likely be galaxies), including AGNs up to higher redshifts.

Two radio continuum images from the ASKAP survey in the direction of the SMC were taken as part of the EMU Early Science Project (ESP) survey of the Magellanic Clouds \citep{Joseph2019} and were observed at 960 MHz and 1320 MHz, respectively. The two source lists that were produced from these images by \cite{Joseph2019} contain radio continuum sources observed at 960 MHz (4489 sources) and 1320 MHz (5954 sources) with a bandwidth of 192 MHz and beam sizes of $30^{\prime\prime}\times30^{\prime\prime}$ and $16\rlap{.}^{\prime\prime}3\times15\rlap{.}^{\prime\prime}1$, respectively. The median RMS noise values were 186 $\mu$Jy beam$^{-1}$ (960 MHz) and 165 $\mu$Jy beam$^{-1}$ (1320 MHz). The observations of the SMC were made with only 33 per cent and 44 per cent (for 960 MHz and 1320 MHz, respectively) of the full ASKAP antenna configuration and 66 per cent of the final bandwidth that was available in the final array, with which the LMC was observed.

The LMC was observed at 888 MHz with a bandwidth of 288 MHz using ASKAP's full array of 36 antennas (scheduling block 8532). The LMC was observed as part of the ASKAP commissioning and early science (ACES, project code AS033) verification \citep{DeBoer2009,Hotan2014,McConnell2016} in order to investigate issues that were found in higher-frequency higher-spectral-resolution Galactic-ASKAP \citep[GASKAP;][]{Dickey2013} survey observations, as well as to test the rapid processing with ASKAPsoft \citep{Whiting2020}. The observations cover a ﬁeld of view of 120 deg$^{2}$, with a total exposure time of 12h40m and a beam size of $13\rlap{.}^{\prime\prime}9\times12\rlap{.}^{\prime\prime}1$. From this image a catalogue of 54,612 sources was created \citep{2021MNRAS.506.3540P}, reaching down to $<$0.2 mJy. 

\begin{figure*}
\centering
\begin{tabular}{c}
	\includegraphics[width=0.95\textwidth]{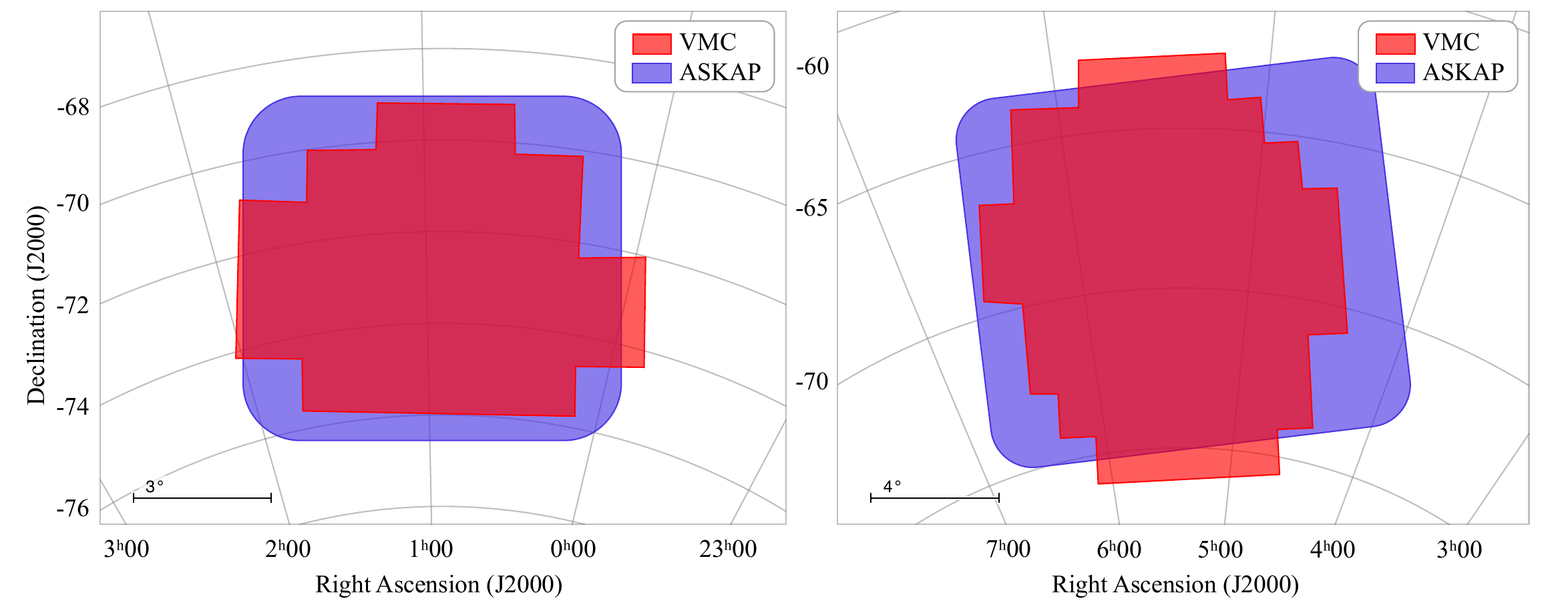}
    \end{tabular}
    \caption{Area of the sky covered by the VMC (near-IR, \textit{red}) and EMU ASKAP (radio, \textit{blue}) surveys of the SMC (left) and LMC (right). }
    \label{fig:MCtsnecoverage}
\end{figure*}

Figure \ref{fig:MCtsnecoverage} shows the comparisons between the area covered by the VMC and EMU ASKAP surveys of the SMC and LMC. 

\subsubsection{All-sky surveys}

The \textit{Gaia} mission \citep{Gaiamission} telescope has the aim of measuring the 3D spatial and velocity distribution of stars, as well as to determine their astrophysical properties. The \textit{Gaia} on-board system is designed to detect point-like sources, but can detect extended sources \citep{GaiaEx} if their central region is sufficiently bright and compact. The latest data release, DR3 \citep{GaiaDR3}, provides celestial positions, proper motions, parallaxes, and broad band photometry in the wide G (centred on 650 nm), blue-enhanced G$_{BP}$ (centred on 360 nm), and red-enhanced G$_{RP}$ (centred on 750 nm) pass-bands. This data release also includes class probabilities (QSO, galaxy or stellar source) for 1.5 billion sources. We use the proper motions in RA and DEC, as well as the three broad band photometry detections from Gaia DR3 as feautures to be used for the machine learning algorithm.

WISE \citep{WISE} is a telescope launched in 2009 to repeatedly map the entire sky in four bands W1, W2, W3, W4 centred at 3.4 $\mu$m, 4.6 $\mu$m, 12 $\mu$m, and 22 $\mu$m, respectively, using a 40-cm telescope feeding arrays with a total of four million pixels. The sensitivities of W1, W2, W3 and W4 correspond to Vega magnitudes of 16.5, 15.5, 11.2, and 7.9, respectively, in the all-sky WISE survey. The AllWISE \citep{Cutri2013} programme extended this work by combining W1 and W2 data from the cryogenic and post-cryogenic survey phases to form the most comprehensive time-lapse view of the mid-infrared sky available. 
We use the W1, W2, W3 and W4 bands from the AllWISE catalogue as features for the machine learning algorithm.

\subsection{Creating the multi-wavelength dataset}

Photometric data can often be missing in some bands, due to, for example, being too faint to be observed at a particular wavelength, or due to a poor observation (from image saturation due to a bright source, artifacts, poor seeing, etc.). However, most machine learning algorithms cannot handle missing data. There are various methods or dealing with missing values, such as removing all rows (where each row would be an individual source and its measurements) that contain at least one missing value, setting the missing values to a value far from the range of real values, or another is to use a model/machine learning algorithm to estimate the missing value from the rest of the dataset. Here, the option of having a "clean" dataset with no missing data was chosen, so that the missing data does not end up affecting the clustering. 

However, cross-matching pre-made catalogues brings about a variety of biases to the final sample. For example, all catalogues have their own flux limits and requiring an detection in all bands skews the sample towards objects that are luminous in all bands. For example, the AGN population of the sample would be biased towards bright AGN seen mostly pole-on (due to requiring an optical detection, which would indicate we are detecting the accretion disk), and excluding the fainter and/or more obscured AGN. Observing resolutions vary across the spectrum, for instance VMC would have sub-arcsecond resolution compared to the beam sizes of several arcseconds in the radio, leading to several resolved near-IR sources falling into a single radio detection. This means nearest-neighbour matching has the potential to misidentify the true counterparts. These biases can be mitigated by using Bayesian methods \citep[e.g. NWAY;][]{NWAY} that calculate the probability of a match based on various factors (e.g. positional errors, colours, etc.), or by using forced photometry \citep[e.g. SExtractor;][]{SExtractor} to measure flux at the exact coordinates of a known source. We expect most of the false matches to arise from the cross-matching of the radio detections to the near-IR detections. In this investigation we choose to see how the machine learning handles these possible false matches and determine if the result can be used as a prior on whether the radio match was true or not.

All coordinate matchings were made to the VMC coordinates, which is an average position taken from the three bands. Cross-matchings between the VMC catalogue and all but ASKAP were done with a 1$^{\prime\prime}$ search radius (as positional accuracy of all these surveys is $<$1$^{\prime\prime}$), keeping only the closest match between the VMC sources and the other catalogues. The cross-matching with ASKAP was done with a 5$^{\prime\prime}$ search radius instead to account for its lower resolution. From cross-matching the individual catalogues we get a distribution of angular separations from each match, these cross-matching radii are chosen such that the expected peak of angular separation and the minimum after it is within the search radius and that the expected increase in matches due to random matches is outside the search radius. Through the use of T\textsc{OPCAT} \citep{Taylor2005} the VMC catalogue was matched with Gaia DR3 \citep{GaiaDR3}, AllWISE and ASKAP radio surveys \citep{Joseph2019,2021MNRAS.506.3540P}. The list of observations adopted as features can be seen in Table \ref{table:PRFparam}. After all the catalogues were matched to the VMC catalogue, the sources with missing data
were removed, the majority of which were due to not having an ASKAP radio detection (7735 and 43,190 of all VMC sources had a corresponding ASKAP detection in the SMC and LMC, respectively). Note that in the Gaia DR3 catalogue, extended sources, such as galaxies which are not expected to have discernable movement across the sky, are given no proper-motion measurements. Therefore, where a source has measured Gaia DR3 photometry, but no proper-motion measurement, proper motion is set to 0 in order to keep radio detected galaxies in the sample. Also note that the AllWISE catalogue includes upper limits on detections (indicated by the photometric quality flag, 'qph', which provides a flag per band), which we have decided to retain in order to keep as many sources as possible, as a radio detected source with a good IR detection in at least one band has a strong likelihood of being extragalactic. This gave 1356 and 6652 radio detected sources ($\sim$17.5\% and $\sim$15.4\% of VMC detected sources with a radio counterpart) across the SMC and LMC, respectively.

\begin{table}
\caption{Observations from multiple surveys are used as features in the t-SNE algorithm. From the optical to infrared photometry new features were created by subtracting each feature from all the other features of the same survey to create colours. Also included as features are the sharpness (a measure of the difference between the observed width of the object and the width of the PSF model, where stars should have a sharpness value of $~$0 and resolved objects sharpness values of $>$0. Sharpness values $<$ 0 indicate artefacts such as bad pixels or cosmic ray impacts.) and proper motions in RA and DEC (pmRA and pmDEC, respectively).}
\vspace{5pt}
\begin{tabular}{|l|l|l|}
        \hline\hline
        Feature & Units & Survey \\
        \hline
        $Y$ PSF & mags (Vega) & VMC \\
        $J$ PSF & mags (Vega) & VMC \\
        $K$\textsubscript{s} PSF & mags (Vega) & VMC \\
        $Y$ sharp PSF & -- & VMC \\
        $J$ sharp PSF & -- & VMC \\
        $K$\textsubscript{s} sharp PSF & -- & VMC \\
        pmRA & mas/yr & Gaia DR3 \\
        pmDEC & mas/yr & Gaia DR3 \\
        $G$ & mags (Vega) & Gaia DR3 \\
        $G$\textsubscript{BP} & mags (Vega) & Gaia DR3 \\
        $G$\textsubscript{RP} & mags (Vega) & Gaia DR3 \\
        W1 & mags (Vega) & AllWISE \\
        W2 & mags (Vega) & AllWISE \\
        W3 & mags (Vega) & AllWISE \\
        W4 & mags (Vega) & AllWISE \\
        888/960/1320 MHz & mJy & ASKAP \\
        \hline
\end{tabular}
\label{table:PRFparam}
\end{table}

Due to the large difference in resolution between the VMC and ASKAP surveys we expect there to be some spurious (false-positive) cross-matches between the two catalogues. The expected fraction of radio detections that have been matched with spurious background sources in the VMC catalogue can be calculated as $f_{spurious}=N_{spurious}/N_{radio}=\sum\pi r^2 \rho_{\rm VMC}$, where the number of spurious cross-matches is $N_{spurious}=N_{radio}\sum\pi r^2 \rho_{\rm VMC}$, $N_{radio}$ is the number of radio sources in the overlapping VMC and ASKAP area, $r$ is the random draw of expected offset between measured radio position and its true counterpart and $\rho_{\rm VMC}$ is the average source density of the VMC field. From cross-matching VMC with the ASKAP surveys we find that the peak of the distribution of separation between the near-IR coordinates and radio coordinates to be $\sim$1.3$^{\prime\prime}$ and $\sim$1$^{\prime\prime}$ for the SMC and LMC, respectively, which we take to reflect the average positional uncertainty of the ASKAP coordinates. From this we find approximate values of the expected spurious cross-match fraction of the radio sources to be  41\% and 46\% for the SMC and LMC, respectively. Note that this simple calculation does not take into account the change in source density across the field, so we expect there to be more spurious cross-matches towards the centre of the Clouds compared to the outskirts.

Optical Survey of the Magellanic Stellar History \citep[SMASH;][]{Nidever2017}, Spitzer SAGE \citep{meixner2006, gordon2011, SAGEdat}, Herschel HERITAGE \citep{meixner2010} and XMM-Newton \citep{2013XMM} X-ray imaging surveys were also considered, but they either do not cover the entirety of the VMC footprint or lack depth in certain wavelength bands, which would cause a lot of missing data and therefore greatly reduce the number of sources in the final catalogue.

Other parameters/features that were added were optical and IR colours, by subtracting each feature from all the other features of the same survey to
create them (e.g. $J-K{\rm s}$ from VISTA, W1$-$W2 from AllWISE, etc.).

\subsection{Catalogue of known sources}
In order to test how well the machine learning is working a catalogue of sources with known classifications is required. We used the compilation of spectroscopically observed sources as such a training set in \citet{Pennock2025}. This provides sources classed as AGN, galaxies, OB stars, RGB stars, RSG stars, AGB stars, post-RGB stars, post-AGB stars, PNe, young stellar objects (YSOs) and foreground stars. To supplement this, we also used SIMBAD to provide examples of rarer classes \citep{Simbad}, such as BLLac/blazars, blue supper giants, white dwarfs, Wolf Rayet stars and various variable classes of stars.

%%%%%%%%%%%%%%%%%%%%%%%%%%%%%%%%%%%%%%%%%%%%%%%%%%%%%%%%%%%%%%%
\section{t-distributed stochastic neighbourhood embedding}\label{sec:tSNE}

Clustering algorithms group objects in a dataset, such that objects in the same group/cluster are more similar to each other than to objects in other groups. Dimensionality reduction algorithms, such as principal component analysis \citep[PCA;][]{jolliffe2002pca}, reduce the number of features in the original dataset, either by selecting a subset of the features that best describe the dataset, or by constructing a new set of features that provide a good description of the dataset. Both clustering and dimensionality reduction algorithms can be used to visualise and interpret complex high-dimensional datasets and uncover hidden trends and patterns in large datasets \citep[e.g.][]{2016ApJS..225...31L,2018A&A...619A.125A,2018Reis,2020ApJ...905...97Z}.

t-SNE \citep[t-distributed stochastic neighbour embedding;][]{2008tsne} is an unsupervised machine learning dimensionality reduction algorithm. It can visualise any high-dimensional dataset by projecting each data-point onto a low-dimensional map, which reveals local as well as global structure of the data at many different scales. t-SNE has been shown to be adept at separating sources into different classes with no prior information about the source nature \citep[e.g.][]{2020ApJ...891..136S}. There is another, similar, algorithm called Uniform Manifold Approximation and Projection \citep[UMAP][]{UMAP}. Compared to t-SNE, UMAP can handle processing larger datasets (tens of thousands to millions) much quicker (minutes compared to hours) and can preserve global structure as well as local structure. However, since this is a continuation of \citet{2022MNRAS.515.6046P} where t-SNE was implemented on the SMC, and we are limiting this work to the radio sample (less than ten thousand sources) and are focused on the more local structure of separating different populations from each other, we decided to continue using t-SNE for the LMC. In this work we use the implementation of t-SNE from the Scikit-learn \citep{scikit-learn} python library.

t-SNE is a non-linear (can capture complex, twisting and crowded structures) and non-deterministic technique (output is slightly different each time it is used on the same data as there is an element of randomness in the optimisation process) using hyperparameters (perplexity, early exaggeration, learning rate and number of steps) with a primary goal to create clusters that can be easily visualised in 2D or 3D. It embeds/projects the points from a higher dimension into a lower dimension whilst trying to maintain the neighbourhood of that point, preserving the local structure of the data. More specifically, the t-SNE technique minimises the divergence (using a cost function called the Kullback-Leibler divergence) between a probability distribution that measures pairwise similarities of the high-dimensional data and a probability distribution that measures pairwise similarities of the low-dimensional points in the embedding. It does this by randomly dropping the data points into the low-dimensional space and then using gradient descent to iteratively nudge similar points together and dissimilar points apart. Unlike the linear Principal Component Analysis (PCA) algorithm, which preserves global structure, t-SNE preserves the local structure, allowing fine structures to be found. In this work we reduce the dimensions down to two, so that the results can be interpreted in a single plot.

The perplexity parameter of t-SNE is related to the number of nearest neighbours that is used in the algorithm, where larger datasets usually require a larger perplexity.
A low perplexity value leads to too much separation between sources (every source is its own cluster), whilst a large perplexity leads to little separation between sources (all sources are part of one cluster). A value of around perplexity = 50 was chosen as it showed well-structured clustering \citep[for examples of the clusterings generated by different clusterings see Figure 3.12 in][]{pennock_thesis}. Note that while points close together can be assumed to be very similar, and points/clusters that are far apart can be assumed to be different, the exact distance between them does not correlate to how different they are.

\section{Results}\label{sec:results}
In this section we present the results of the t-SNE algorithm on the optical to radio data of the LMC and SMC as 2D maps, as shown in Fig. \ref{fig:MCtsneclasses}, where the SMC t-SNE map is the same as used in \citet{2022MNRAS.515.6046P}. We show, approximately, how various classes of sources are clustered across the maps from the analysis that follows and from previously classified sources from the literature, as well as how some of the features used vary across the map and align with the various classes.

\subsection{t-SNE maps}

\begin{figure*}
\centering
\begin{tabular}{c}
	\includegraphics[width=0.95\textwidth]{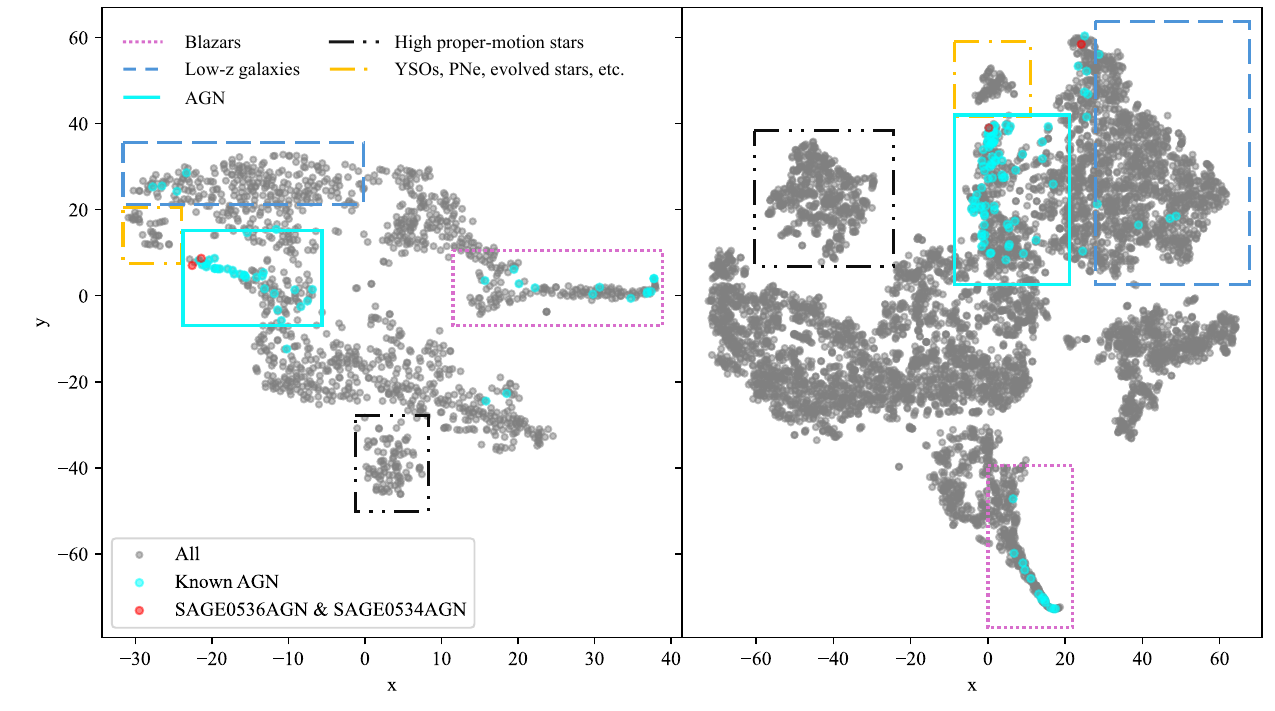}
    \end{tabular}
    \caption{t-SNE map of the SMC (left) and LMC (right). Known, spectroscopically observed AGN are highlighted in cyan and SAGE0536AGN and SAGE0534AGN, that have been previously mistaken for dusty evolved stars, have been highlighted in red. The rectangular regions show the general area of specific classifications of sources. Note that the dusty AGN have been separated in the LMC plot, whereas they are together in the SMC plot.}
    \label{fig:MCtsneclasses}
\end{figure*}

Using known classifications from the literature we can loosely classify the sources of the t-SNE maps of the SMC and LMC -- see Figure \ref{fig:MCtsneclasses}. To do this we used classifications from SIMBAD \citep{Simbad} and from the training set used for supervised machine learning in \citet{Pennock2025}, cross-matched within 1$^{\prime\prime}$. 

High proper-motion ($5\lesssim$ PM $\lesssim 90$ milli-arcsec/yr) stars are outlined in a black dashed-dot-dot line and are an isolated group of sources in both maps. These are all most likely stellar sources in front of the true radio sources. This result shows that this technique could be a good way to remove serendipitous false cross-matches between radio surveys and shorter wavelength photometry surveys. Some of these, however, could possibly be one of the rare radio active stars \citep[e.g.][]{2022LMCpulsar}. Another option is that these are nearby galaxies that have been given proper motions due to variable sources \citep[e.g][]{2017MNRAS.471.3775P,2023A&A...674A..11D}. For example, an SN could occur in a galaxy and alter the centre of brightness for the source, or an AGN becoming more or less luminous than its host galaxy. Nearby galaxies are also extended sources, but the calculation of proper motions is designed for point sources, so therefore applying this to an extended source has potential to lead towards miscalculated values of proper motion.

In the regions outlined by a yellow/orange dashed-dot line, there is a small island of YSOs, PNe, evolved/dusty stars, emission-line stars, etc, which are sources that are often misclassified as AGN and vice versa. In \citet{2022MNRAS.515.6046P}, there were two spectroscopically observed sources, from an area of transitioning galaxies hosting AGN on the SMC t-SNE map, and whose true classifications turned out to be emission-line and evolved dusty stars. These sources were found near the red points on the SMC t-SNE map, closest to the small island of YSOs, PNe, evolved/dusty stars, emission-line stars, etc. If some of the stellar sources can be found on the outskirts of the AGN section of the map, this could imply the opposite, that some transitioning galaxies hosting AGN are on the edge of the small stellar island closest to the sample investigated in this work.

AGN that have been spectroscopically observed can be found roughly along the centre of the map, y $\sim$ 0 for the SMC and x $\sim$ 0 for the LMC, and are outlined by a solid cyan line. Blazars can be found to the far right of the AGN sequence in the SMC and to the far bottom of the AGN sequence of the LMC map, and are outlined by a magenta dotted line. See Section \ref{blazseq} for discussion on spectroscopic follow-up on sources in this blazar candidate sequence.

Low-redshift (low-$z$, where $z\lesssim$0.8) galaxies, with obvious galaxy shape in survey images, can be found from the AGN region towards the top of the t-SNE map in the SMC, and to the far right in the LMC, and are outlined in a dashed blue line. There is a redshift gradient in both maps with $z$ increasing  towards the AGN sequence (down and to the left in the SMC and LMC, respectively).

The LMC map, unlike the SMC map, separates SAGE0534AGN from SAGE0536AGN (see red circles Figure \ref{fig:MCtsneclasses}). These two sources can be found on two separate points that reach towards the island of YSOs, PNe, evolved stars and emission line stars, which is similar to the SMC map. SAGE0534AGN can be found among the AGN sequence and SAGE0536AGN can be found among the low-$z$ galaxies, which makes sense with respect to the redshift gradient in the map (see Section \ref{sec_redshift} and Figure \ref{fig:z}) and that the host galaxy is detected for SAGE0536AGN but not for SAGE0534AGN \citep{2022MNRAS.515.6046P}.

The co-ordinates of the SMC and LMC sample investigated in this work along with the two dimensions that resulted from the t-SNE algorithm, $x$ and $y$ (e.g. as seen in Figure \ref{fig:MCtsneclasses}), and the PRF classifications and their probabilities from \citet{Pennock2025}, as mentioned in Section \ref{sec:PRF}, will be made available as supplementary material, and the first 5 rows of the table can be seen in Table \ref{tab:results}.

\begin{table*}
    \centering
    \caption{First 5 rows of the results table provided as supplementary material alongside the paper. For each source it provides the field the source can be found in (SMC or LMC), the $x$ and $y$ values provided by the t-SNE algorithm, the Right Ascension and Declination of the object from the VMC catalogue in degrees, and the classification ('PRF\_Class') and its probability (PRF\_Class Prob.) from \citet{Pennock2025}, as mentioned in Section \ref{sec:PRF}.}
    \begin{tabular}{ccccccc}
        \hline\hline
         Field & t-SNE\_x & t-SNE\_y  & VMC\_RA  & VMC\_DEC  & PRF\_Class & PRF\_Class Prob. \\
         &  &  & (J2000) deg. & (J2000) deg. &  & \\
         \hline
         LMC & -49.28 & -9.17 & 65.60571 & -70.62935 & Unk & 0.326 \\
         LMC & 11.21 & -5.92 & 65.60571 & -70.66553 & AGN & 0.903 \\
         LMC & 40.98 & 18.73 & 65.76690 & -70.67836 & Galaxy & 0.806 \\
         LMC & 28.20 & 10.54 & 65.96743 & -70.54397 & Galaxy & 0.617\\
         LMC & 27.65 & 15.38 & 66.04878 & -70.51840 & Galaxy & 0.561\\
         ... & ... & ... & ... & ... & ... & ... \\
         \hline
    \end{tabular}
    \label{tab:results}
\end{table*}

\subsection{Feature distribution}

\begin{figure*}
\centering
\begin{tabular}{c}
	\includegraphics[width=\textwidth]{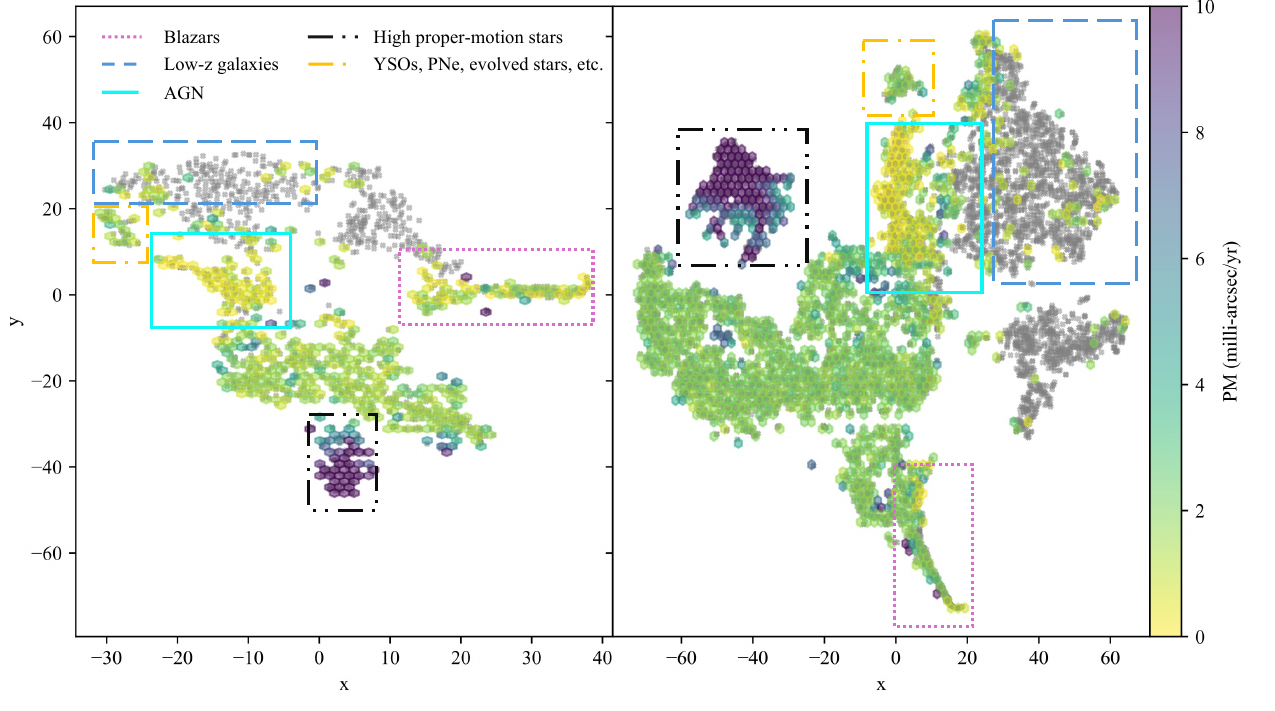}
    \end{tabular}
    \caption{Comparison of the distribution of proper motions of the sources in the t-SNE maps of the SMC (left) and LMC (right). The entire sample is plotted as grey points. On top of this we have colour-coded the space by the median proper motion (PM) per hexagon. The rectangular regions are the same as in Figure \ref{fig:MCtsneclasses}. As expected, the highest PM values are where high proper-motion stars are located, and the lowest PM values are where AGN are located. It should be noted that Gaia does not provide PM values for likely extragalactic (extended) sources, which is why most of the sources in the galaxy area have no PM measurements and are represented as grey points. As extended galaxies are the likely counterparts for the radio detections (e.g. AGN, star-formation activity), we set their PM measurements to zero to allow these sources to be accepted by the t-SNE algorithm, but decided not to plot them here as we do not know their true measurements.}
    \label{tsnemapsPM}
\end{figure*}

The Gaia DR3 catalogue provides proper-motions measurements, and from the distribution of the total proper motions across the t-SNE maps, as shown in Figure \ref{tsnemapsPM} (note that we use the proper motion in RA and DEC as features for the t-SNE algorithm, as opposed to the total proper motion which is a combination of the two), it can be seen that most sources show Magellanic proper motions \citep[e.g.][]{Cioni2016} or lower. The lowest proper motions can be matched with known AGN and galaxies, as expected. Where there are no data for the proper motions, which we set to 0 so that it could be used with t-SNE (they are not plotted here as we do not know their true measurements), is where most of the galaxies can be found. 

High proper-motion (foreground/Milky-Way) stars can be found in the isolated bottom and top left islands of the SMC and LMC t-SNE maps, respectively. These sources have the potential to be radio stars. Further inspection of these optical/IR sources in the radio have shown that they are a mixed bag. Some of the sources match up with artefacts around bright radio sources that have made their way into the radio catalogues, some are so faint they could be part of the background variation and others have matched up with jets/lobes/extended emission far from the host of the radio sources. Some do match up with radio point sources, so could be foreground radio stars. Indeed, foreground cool dwarf systems with radio emission were identified in the LMC ASKAP image \citep[see][]{2021MNRAS.506.3540P, 2023Rose}. It is possible there exists a large population of radio stars just below or around the sensitivity of the ASKAP survey.

\begin{figure*}
\centering
\begin{tabular}{c}
	\includegraphics[width=\textwidth]{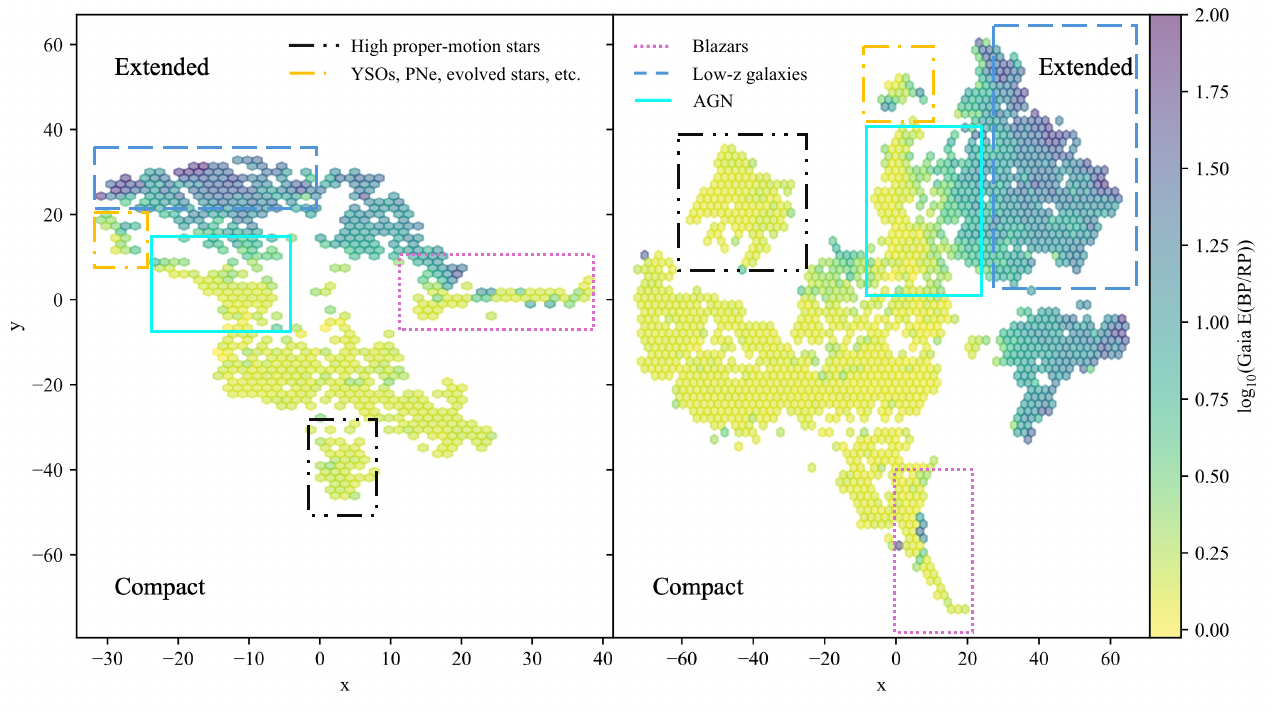}
    \end{tabular}
    \caption{Comparison of the distribution of the median value of Gaia BP/RP excess of the sources in the t-SNE maps of the SMC (left) and LMC (right). Large values of Gaia BP/RP excess indicate a large difference between $G$ flux the average of the integrated flux values of $G_{\rm RP}$ and $G_{\rm BP}$, which for a star-like objects would be the same, but for extended objects would be different. The rectangular regions are the same as in Figure \ref{fig:MCtsneclasses}.}
    \label{tsnemapsGaiacol}
\end{figure*}

The Gaia DR3 catalogue provides measurement known as the Gaia BP/RP excess factor, a measure of the difference between the sum of the integrated $G_{\rm RP}$ and $G_{\rm BP}$ fluxes and the $G$ flux, where the further from 0 the Gaia BP/RP excess factor, Gaia E(BP/RP) \citep{DeAngeli2023}, deviates the more inconsistent the $G_{\rm RP}$ and $G_{\rm BP}$ fluxes are with the $G$ flux. This factor identifies extended objects because the value of $G$ for a point source is expected to roughly equal the average of $G_{\rm RP}$ and $G_{\rm BP}$, which can be seen is true for areas where known stars and higher redshift sources are found. The larger magnitude differences are caused by the way the flux is measured for $G$ (broad pass band filter) compared to $G_{\rm RP}$ and $G_{\rm BP}$ (prism and colour filter). In Figure \ref{tsnemapsGaiacol} we show how Gaia E(BP/RP) varies across the t-SNE maps. From this we can see that galaxies can be found where the greatest difference between Gaia $G$ and the $G_{\rm RP}$ and $G_{\rm BP}$ bands, as expected.

\begin{figure*}
\centering
\begin{tabular}{c}
	\includegraphics[width=\textwidth]{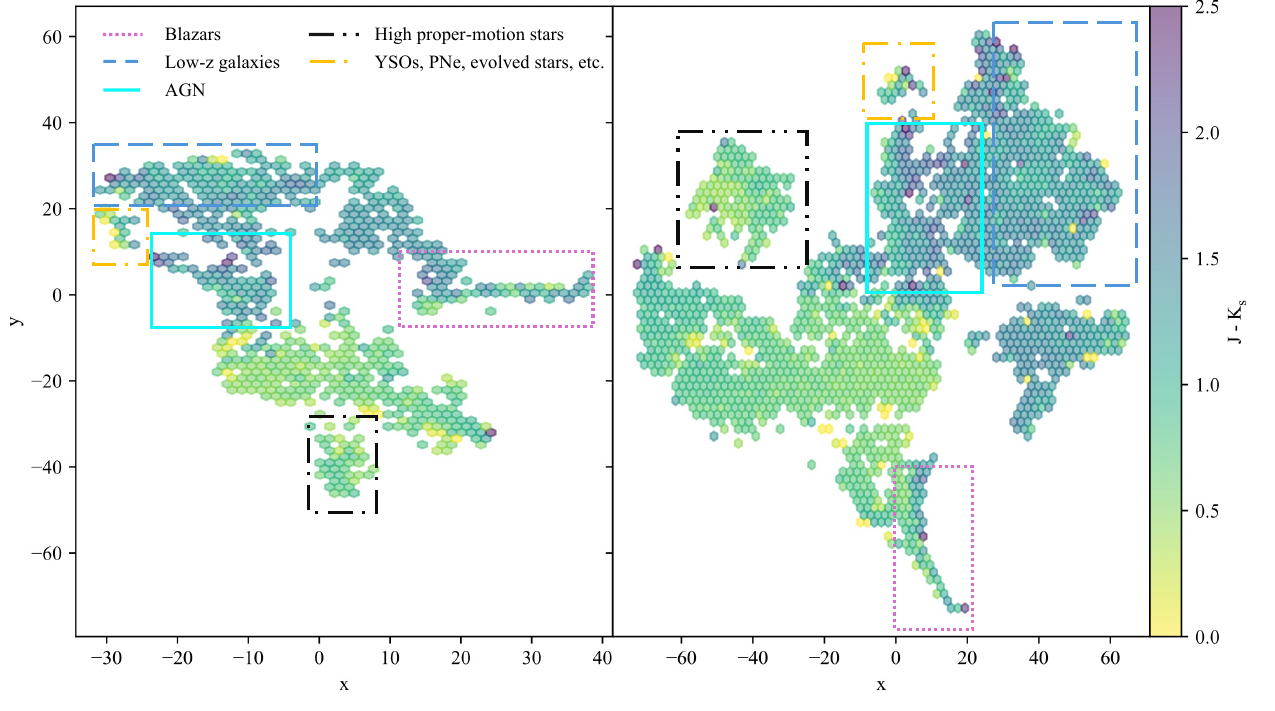}
    \end{tabular}
    \caption{Comparison of the distribution of the median value of VMC PSF colour $J- K_{\rm s}$ (mag) of the sources in the t-SNE maps of the SMC (left) and LMC (right). The rectangular regions are the same as in Figure \ref{fig:MCtsneclasses}. It can be seen that the extragalactic regions of the map have redder $J- K_{\rm s}$ colours.}
    \label{tsnemapsVMCcol}
\end{figure*}

\begin{figure*}
\centering
\begin{tabular}{c}
	\includegraphics[width=\textwidth]{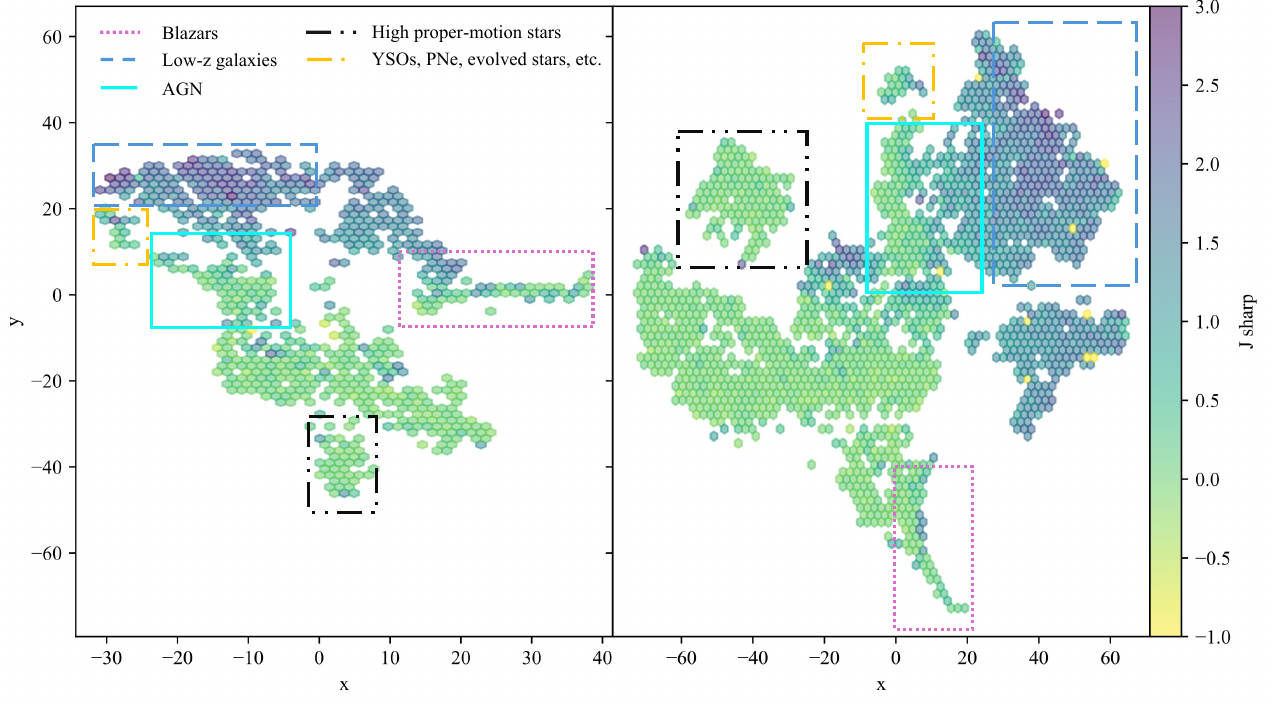}
    \end{tabular}
    \caption{Comparison of the distribution of the median value of VMC PSF J-band sharpness value of the sources in the t-SNE maps of the SMC (left) and LMC (right). The rectangular regions are the same as in Figure \ref{fig:MCtsneclasses}. It can be seen that the more extended objects (J sharp $>$ 0) line up with the low-$z$ galaxies. }
    \label{tsnemapsVMCsharp}
\end{figure*}

The VISTA parameters that show the greatest difference across class positions on the map are the colour $J- K_{\rm s}$ (Fig. \ref{tsnemapsVMCcol}) and the $J$ sharpness parameter (Fig. \ref{tsnemapsVMCsharp}).  The $J- K_{\rm s}$ colour is reddest where the extragalactic sources can be found. The $J$ sharp parameter is highest (least sharp) to the far right where the lowest redshift galaxies reside, which would be extended in the VMC survey. Note that the distributions of both $J- K_{\rm s}$ and the $J$ sharpness are very similar, most likely because redder sources are generally associated with extended sources such as galaxies and star-forming regions. The areas where AGN are found have red $J- K_{\rm s}$ and low $J$ sharpness (point-like), as expected. 

\begin{figure*}
\centering
\begin{tabular}{c}
	\includegraphics[width=\textwidth]{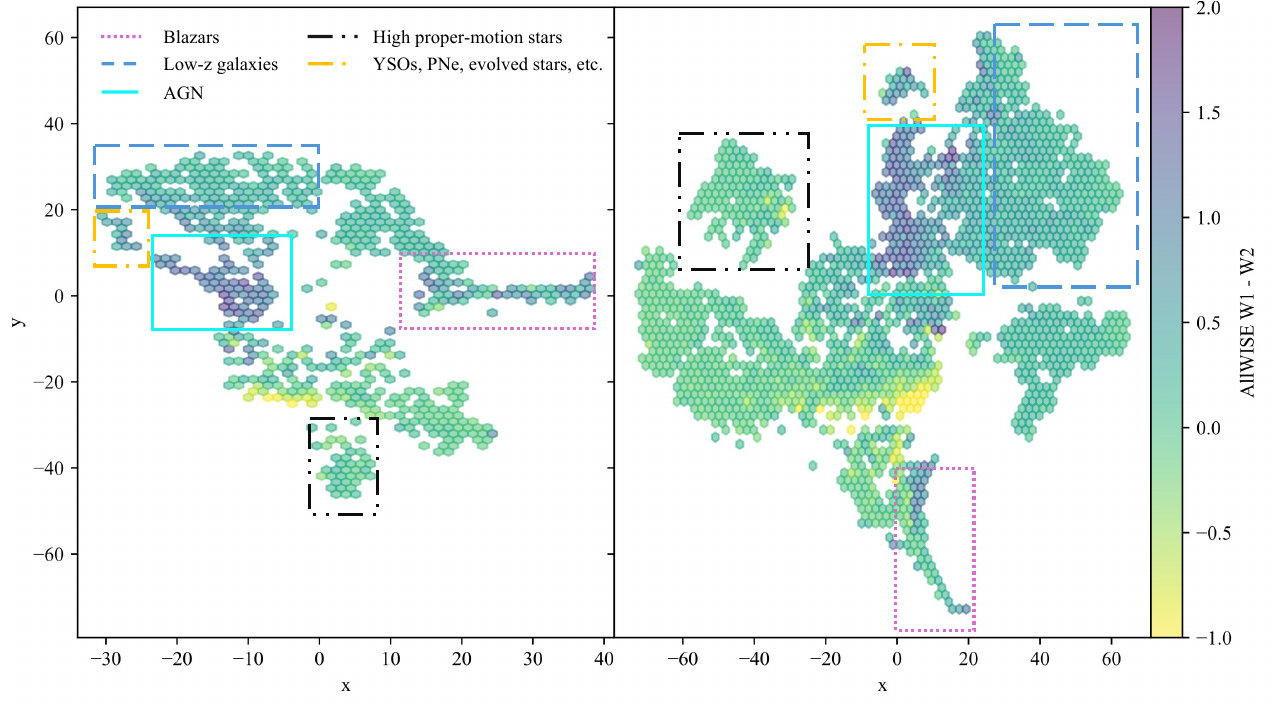}
    \end{tabular}
    \caption{Comparison of the distribution of the median value of AllWISE colour W1$-$W2 of the sources in the t-SNE maps of the SMC (left) and LMC (right). The rectangular regions are the same as in Figure \ref{fig:MCtsneclasses}. Known AGN clearly stand out in this AllWISE colour (being red).}
    \label{tsnemapsWISE}
\end{figure*}

The WISE parameter that shows the greatest difference across class positions on the map is the W1$-$W2 colour (Figure \ref{tsnemapsWISE}). The highest values coincide with the known AGN, showing that this WISE colour is good for selecting AGN, in agreement with works such as \cite{WISE,2012Mateos,Stern2012,Assef2013,Nikutta2014}. Compared to nearby galaxies and star-forming objects, AGN stand out more in W1 $-$ W2 than $J$ $-$ $K$\textsubscript{s}.

In Figure \ref{tsnemapsWISE}, the area highlighted in light green/yellow only stands out in the W1$-$W2 colour, where it is more negative than elsewhere, and W2 mag, where they tend to be fainter than other sources. After inspecting where the AllWISE W2 bands have good detections compared to where upper limits have been used, this area coincides with sources that mostly have upper limits in this band (see Section \ref{sec:AllWISEflags}).

\begin{figure*}
\centering
\begin{tabular}{c}
	\includegraphics[width=\textwidth]{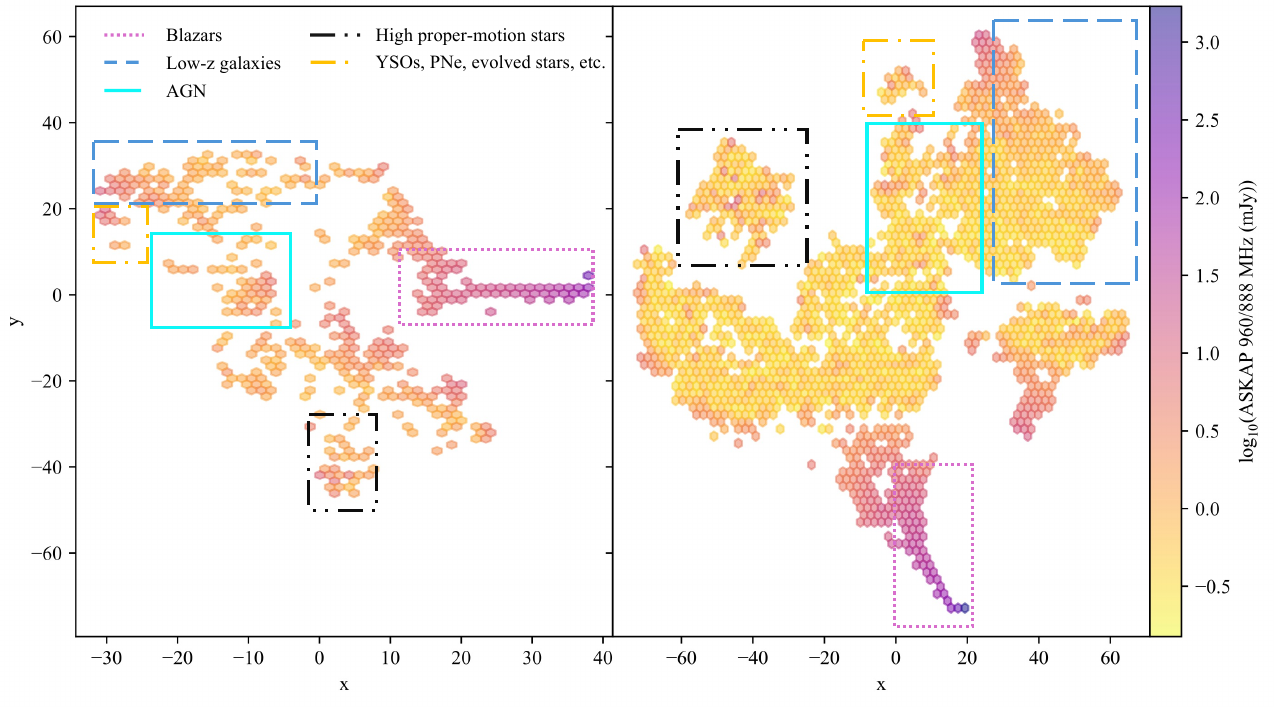}
    \end{tabular}
    \caption{Comparison of the distribution of the median value of ASKAP 960/888 MHz flux density in the t-SNE maps of the SMC (left) and LMC (right). Note that the missing sources in the SMC are those detected in the 1320 MHz image but not the 960 MHz image. The rectangular regions are the same as in Figure \ref{fig:MCtsneclasses}. The area of the brightest radio sources lines up with the area of known blazars, as expected.}
    \label{fig:radio}
\end{figure*}

The majority of radio detections are expected to be extragalactic \citep[e.g.][]{2021MNRAS.506.3540P}. It is mostly due to synchrotron radiation caused by AGN, but can also be caused by star-forming processes at the low luminosity end. Of the stellar sources, H\,\textsc{ii}/YSO and PNe are expected to be associated with free--free radio emission, and foreground stars that are close enough that a radio detection is possible. In Figure \ref{fig:radio} we show the distribution of the flux density in the 960/888~MHz band across the two t-SNE maps. From this we can see that the median value of radio flux density is broadly similar across the maps, except to the far right in the SMC t-SNE map, and to the bottom of the LMC t-SNE map, where the median flux density is much greater. This lines up with the known blazar sources, which are AGN whose relativistic jets are Doppler boosted along our line of sight, causing them to be brighter.

Overall, the LMC and SMC t-SNE maps show similar distributions, which can be seen when rotating the SMC map 90 degrees clock wise and compared to the LMC map. However, the LMC exhibits a greater apparent spread of sources and more complex substructure within several populations, particularly in regions outside the defined classification boxes, where sources may have proper motions consistent with stellar sources belonging to the Magellanic Clouds (see Figure \ref{tsnemapsPM}).These differences could partly reflect the LMC's larger and more metal-rich stellar population compared with the SMC. Higher metallicity can influence stellar evolution, mass loss and circumstellar dust production, resulting in differences in the observed infrared colours and magnitudes of evolved stars \citep[e.g.][]{Boyer2011, 2019Dalal}. Combined with differences in star-formation history, this could produce a broader range of photometric properties among sources associated with the LMC, potentially contributing to the more complex clustering observed in its t-SNE map. Nevertheless, the similarity between the maps suggests that the two fields contain broadly comparable classes of radio-detected sources, which could be in part due to a substantial fraction of the radio-detected sources likely being background galaxies and AGN rather than members of the Magellanic Clouds \citep[e.g.][]{2021MNRAS.506.3540P}.

\section{Discussion}\label{sec:disc}
In this section we compare the t-SNE map clustering to other features that would indicate an AGN presence (Section \ref{extrafeats}), we assess the quality of the AllWISE data and its effect on the t-SNE map (Section \ref{sec:AllWISEflags}), we compare the t-SNE maps to other classification methods used on the Magellanic Clouds in past literature (Section \ref{sec:otherclass}) and we discuss the follow-up spectroscopy of sources from the blazar candidate sequence (Section \ref{blazseq}).

\subsection{Features not used in t-SNE}\label{extrafeats}
Not all available features were used in the creation of the t-SNE map, mainly because incompleteness of these features would restrict the size of the sample even further. Features not included are redshift from spectroscopic observations, variability information from the multi-epoch VMC survey, deeper mid-IR observations from the Spitzer SAGE surveys and X-ray detections. 

\subsubsection{Redshift}\label{sec_redshift}
\begin{figure*}
\centering
\begin{tabular}{c}
	\includegraphics[width=\textwidth]{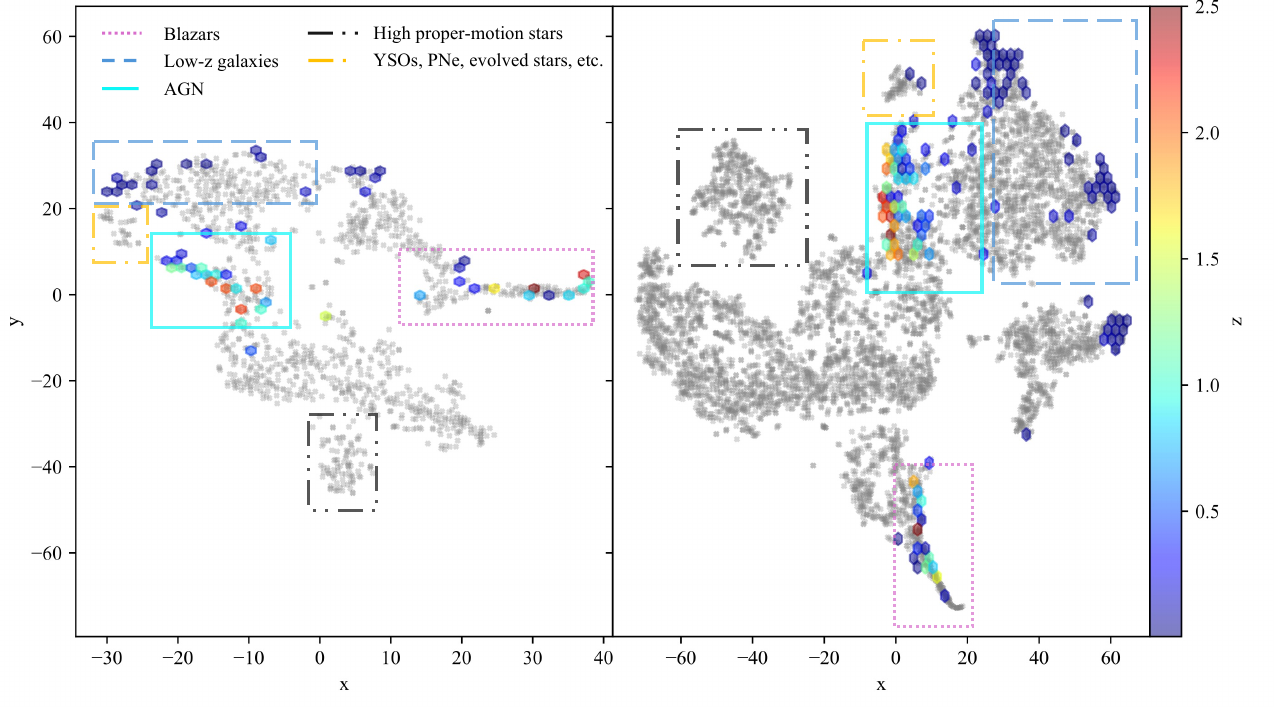}
    \end{tabular}
    \caption{Comparison of the distribution of redshift in the t-SNE maps of the SMC (left) and LMC (right). The rectangular regions are the same as in Figure \ref{fig:MCtsneclasses}. It can be seen that there is a redshift gradient from top to bottom for the SMC and right to left for the LMC. The lowest redshifts match with the most extended sources, the low-$z$ galaxies.}
    \label{fig:z}
\end{figure*}

Redshifts are taken from the compilation of spectroscopically observed sources in \citet{Pennock2025}\footnote{see Table 1 in \citet{Pennock2025} for a summary.} which number 275 and 75 for the LMC and SMC, respectively. Plotting these across the t-SNE maps in Figure \ref{fig:z} reveals a redshift gradient with the lowest redshifts on the top (right) which increase towards the bottom (left) for the SMC (LMC) t-SNE map.

This redshift gradient is most likely picked up by the t-SNE algorithm due to sources becoming redder at higher redshifts, and the VMC sharpness values providing information on how extended a source appears (more extended, lower redshift).

\subsubsection{VISTA Variability}

AGN are known aperiodic variable sources, varying on a wide range of timescales from minutes to years \citep[e.g.][]{padovani2017}. The VMC survey was a multi-epoch survey that started in 2009, observed over $\sim$ 9 years with some additional epochs up to 2023, so it can detect AGN variability. In the VMC catalogues a probability of a source being variable has been calculated, as described in \cite{2009Cross}.

\begin{figure*}
\centering
\begin{tabular}{c}
	\includegraphics[width=\textwidth]{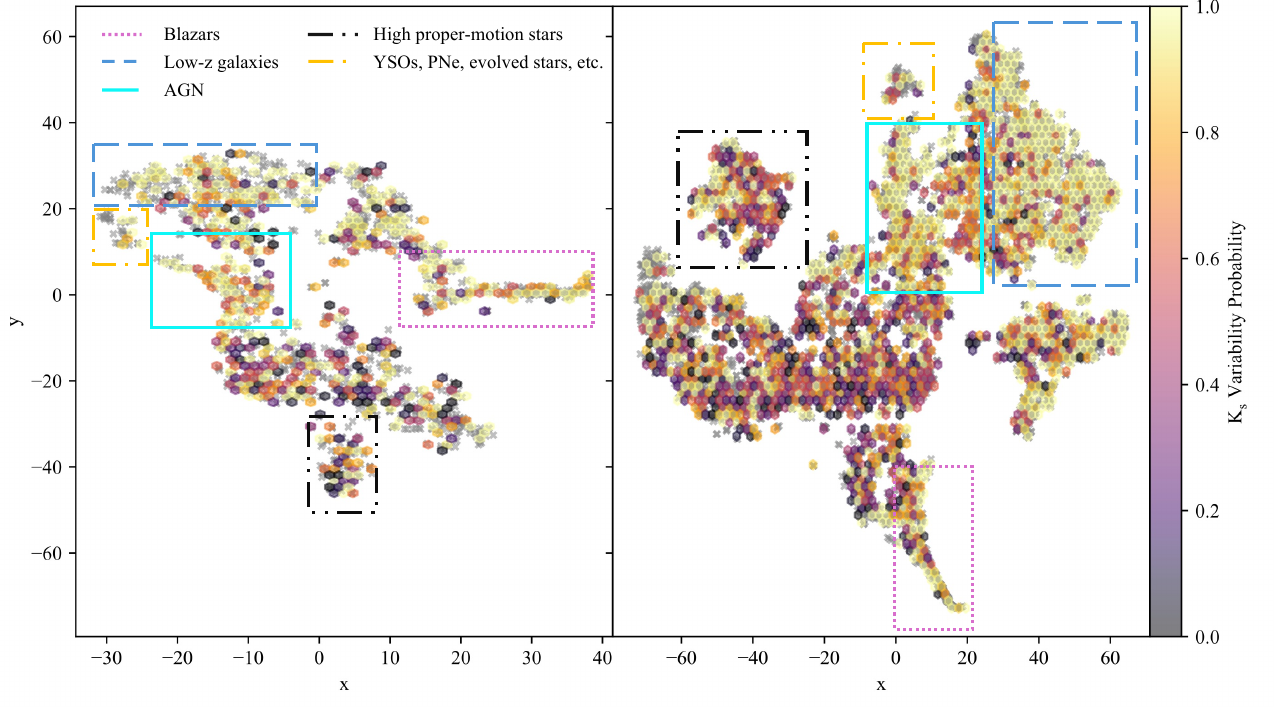}
    \end{tabular}
    \caption{Distribution of the median value of VMC probability of a source being variable in the $K_{\rm s}$ band across the t-SNE maps of the SMC (left) and LMC (right). The rectangular regions are the same as in Figure \ref{fig:MCtsneclasses}. It can be noted that the highest probabilities for a source being variable are found amongst the AGN area.}
    \label{fig:VMCVAR}
\end{figure*}

Figure \ref{fig:VMCVAR} shows how the median value of variability probability in the VISTA $K_{\rm s}$ band varies across the tSNE maps. 
From this we see that in the areas where AGN/blazars reside the variability probability is generally higher (more yellow). We also see in the galaxy wedge, that towards increasing extendedness (higher values of $J$ sharp, see Figure \ref{tsnemapsVMCsharp}) and therefore lower redshift, there is a correlation with variability probability. Whilst this could indicate a population of low luminosity AGN, it  could also be due to the extended profile causing errors in the calculated photometry, and that when selecting AGN using variability this should be taken into account. 

\subsubsection{mid-IR: Spitzer}

\begin{figure*}
\centering
\begin{tabular}{c}
	\includegraphics[width=\textwidth]{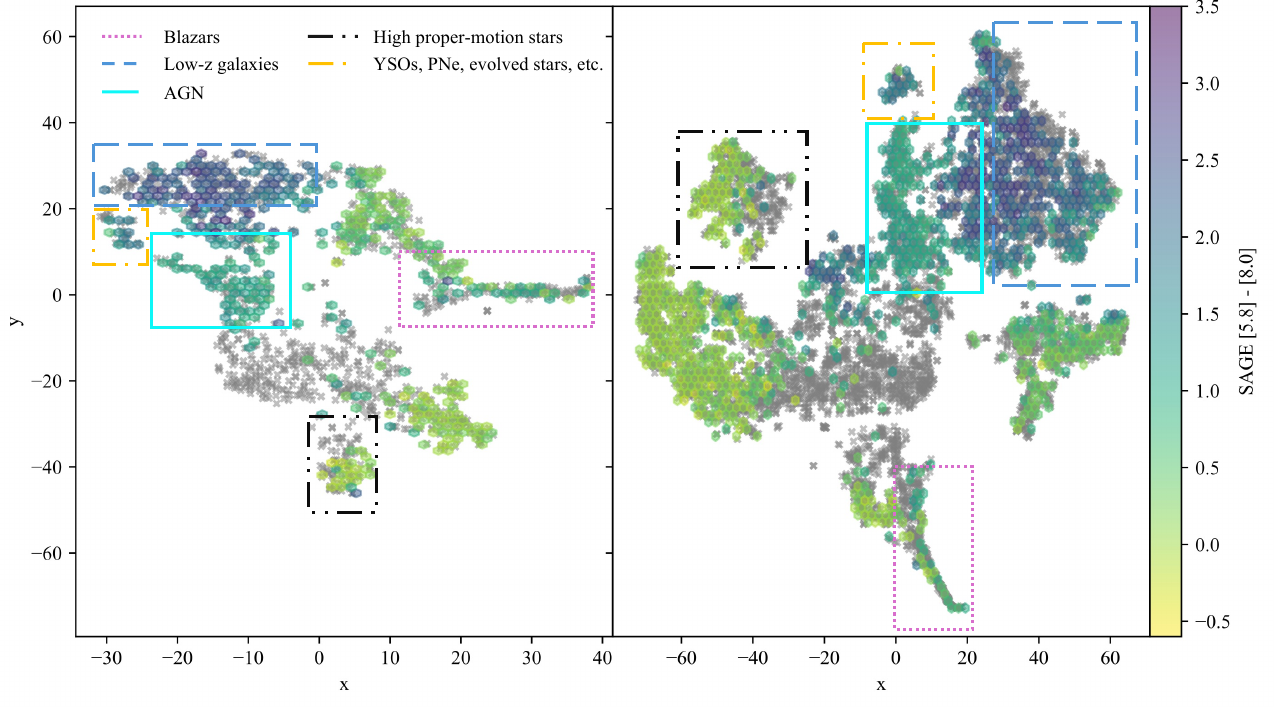}
    \end{tabular}
    \caption{Comparison of the distribution of the median value of Spitzer IRAC [5.8]--[8.0] colours in the t-SNE maps of the SMC (left) and LMC (right). The rectangular regions are the same as in Figure \ref{fig:MCtsneclasses}. This map shows a clear colour difference between the stellar and extragalactic populations, and between the AGN and galaxies.}
    \label{fig:tsneGSAGE8p0}
\end{figure*}

The mid-IR Spitzer SAGE survey was not used due to its smaller field compared to the VMC. Of the bands Spitzer observed in, the 5.8$\mu$m and 8.0$\mu$m bands did not have an analogous waveband in the other surveys (Spitzer IRAC 3.6$\mu$m and 4.5$\mu$m closely align with W1 and W2, respectively) used in the tSNE algorithm. We plot the Spitzer IRAC [5.8]--[8.0] colour across the t-SNE maps in Figure \ref{fig:tsneGSAGE8p0}. The colours can be seen to separate the AGN and galaxies from the stellar sources, though not necessarily from the dusty evolved stellar sources. 

There is a noticeable dearth of sources at around (0,$-$20). In the Spitzer IRAC 3.6 and 4.5$\mu$m (comparable to WISE 1 and 2) there is no obvious gap, and in the W3 and W4 ($\sim$11.5 and 22$\mu$m) the sources are observed. The AllWISE catalogue flags indicated these as upper limits for the W3 and W4 bands (see Section \ref{sec:AllWISEflags}) which would match with these sources not being detected in SAGE 5.8$\mu$m and 8.0$\mu$m. Furthermore, the near-IR and SAGE [3.6]--[4.5] colours are compatible with normal RGB stars, so most likely a mismatch between extragalactic radio sources and Magellanic stars in crowded areas of the MCs.

\subsubsection{X-ray}

The eRASS1 catalogue \citep{2024Merloni} is derived from an all-sky survey performed with the eROSITA (extended ROentgen Survey with an Imaging Telescope Array) instrument onboard the SRG (Spectrum-Roentgen-Gamma) satellite \citep{2021Sunyaev}. It is a wide-field X-ray telescope observing in the energy range $\sim$0.2--8~keV. We cross-matched the coordinates of the t-SNE sources with the to the optical counterparts (from Legacy Survey DR10) in the eRASS DR1 counterpart catalogue \citep{2025Salvato}, with a 1$^{\prime\prime}$ search radius , which yielded 86 and 555 sources in the SMC and LMC, respectively. More recently, the eRASS3 catalogue (which has doubled the X-ray source content of eRASS1) was released alongside an optical counterpart catalogue, eRASS DR2 counterpart catalogue \citep{eROSITA_DR2}. Cross-matching eROSITA DR2 optical counterparts to the t-SNE sources, with a 1$^{\prime\prime}$ search radius, yielded 148 and 841 sources in the SMC and LMC, respectively (note that there are 5 and 45 t-SNE sources in the SMC and LMC, respectively, that have counterparts in the eRASS DR1 catalogue, but not in the eRASS DR2 catalogue).

\begin{figure*}
\centering
\begin{tabular}{c}
	\includegraphics[width=\textwidth]{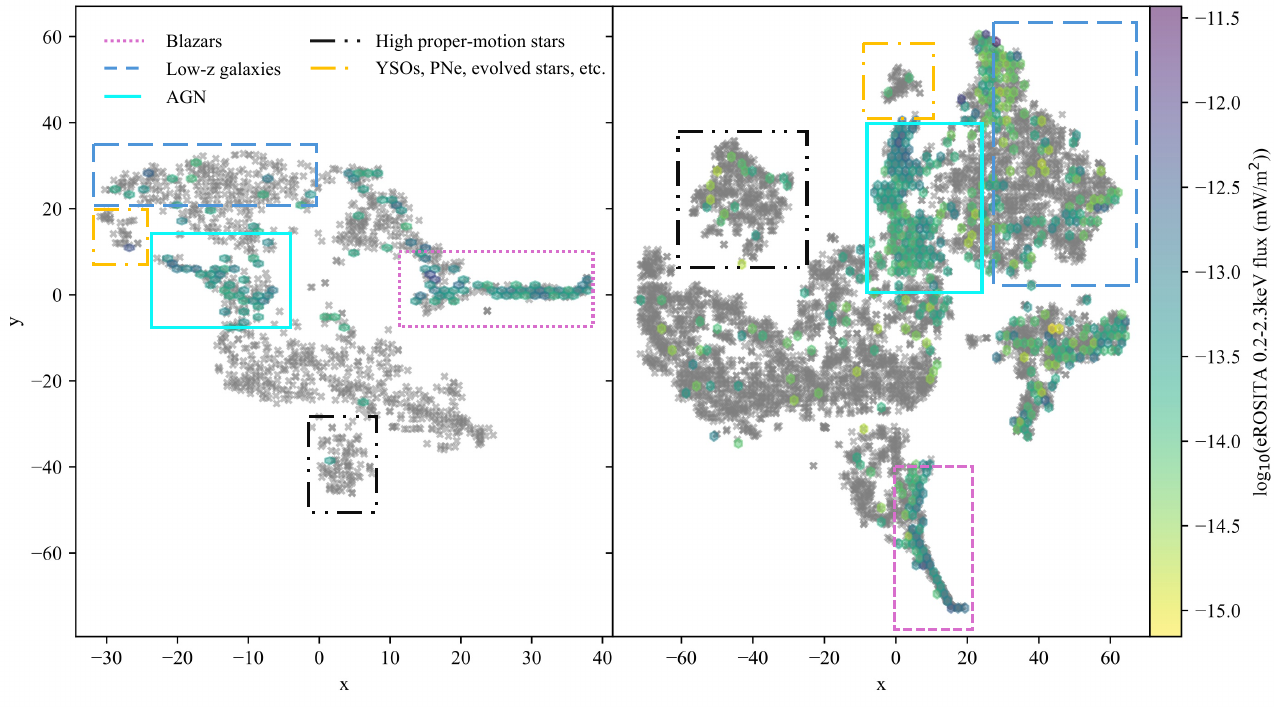}
    \end{tabular}
    \caption{Comparison of the distribution of the median value of eROSITA 0.2--2.3keV fluxes in the t-SNE maps of the SMC (left) and LMC (right). It can be seen that the majority of X-ray detections align with the areas in which AGN and blazars are located.}
    \label{fig:tsneerASS}
\end{figure*}

The majority of X-ray detections are expected to be due to AGN but, towards the Magellanic Clouds, we may observe X-ray binaries and chromospherically active foreground stars as possible stellar originators of X-ray emission. Figure \ref{fig:tsneerASS}, where we plot X-ray detections on the t-SNE maps and their corresponding fluxes, confirms that the highest concentration of X-ray sources is where AGN are found. The low redshift galaxies that have an X-ray detection could be an indication of an obscured AGN population.

We also find that 2 and 19 eRASS DR1 sources (and an additional 1 and 7, respectively, from eRASS DR2) in the SMC and LMC, respectively, that are located in the foreground star area of the t-SNE maps are X-ray detected. We list these X-ray and radio detected foreground (Milky Way) stars candidates in Table \ref{XRstar}, along with their brightness in Gaia $G$ band and their 0.2-2.3keV flux. Furthermore, \citet{2024Freund} cross-matched eRASS DR1 sources with Gaia DR3 in order to find coronally active X-ray stars, and we find that 7/13 of the eRASS DR1 sources have been identified as likely coronally active stars.

\begin{table}
    \centering
    \caption{List of sources in the foreground star region of the SMC and LMC t-SNE maps that are both radio ASKAP detected and X-ray eROSITA sources. The 'Coronal' column indicates whether the source was identified as a highly probable coronally emitting star in \citet{2024Freund}, where 'T' indicates true and '-' indicates that the eROSITA source is in the DR2 catalogue but not the DR1 catalogue, and was therefore not investigated in \citet{2024Freund}.}
    \begin{tabular}{cccccc}
    \hline\hline
        \llap{R}egio\rlap{n} & IAU Name &  Gaia $G$ & 0.2-2.3keV flux & \llap{co}ron\rlap{al?} \\
        & & (mag) & ($10^{-14}$ mW/m$^2$) &  & \\
         \hline
       \llap{S}MC & \llap{3}eRASS J010131.3-74451\rlap{4} &  18.4 & 3.31 & - \\
       \llap{S}MC & \llap{1}eRASS J014111.1-75035\rlap{0} &  18.4 & 3.31 & \\
       \llap{L}MC & \llap{1}eRASS J043530.9-71473\rlap{8} & 17.7 & 2.44  & \\
       \llap{L}MC & \llap{1}eRASS J050108.7-67322\rlap{9} &  15.5 & 1.18   & \\
       \llap{L}MC & \llap{3}eRASS J051032.7-66011\rlap{0} & 17.8 & 0.48 & - \\
       \llap{L}MC & \llap{1}eRASS J051358.6-64121\rlap{7} & 20.4 & 1.03   & \\
       \llap{L}MC & \llap{1}eRASS J052347.5-65272\rlap{2} &  20.2 & 1.32   & \\
       \llap{L}MC & \llap{1}eRASS J054933.5-66284\rlap{9} &  15.5 & 0.48   & T\\
       \llap{L}MC & \llap{3}eRASS J055011.7-66175\rlap{4} & 18.8 & 0.19 & - \\
       \llap{L}MC & \llap{1}eRASS J055118.9-68015\rlap{0} & 17.3 & 1.40   & T\\
       \llap{L}MC & \llap{1}eRASS J060340.8-66393\rlap{7} &  12.7 & 0.33   & T\\
       \llap{L}MC & \llap{1}eRASS J060513.4-69150\rlap{0} &  15.3 & 0.66   & T\\
       \llap{L}MC & \llap{3}eRASS J060602.4-72102\rlap{5} & 15.7 & 0.90 & - \\
       \llap{L}MC & \llap{3}eRASS J060954.7-68303\rlap{3} & 14.5 & 0.26 & - \\
       \llap{L}MC & \llap{3}eRASS J061139.2-67092\rlap{5} & 18.7 & 0.70 & - \\
       \llap{L}MC & \llap{1}eRASS J061519.9-71070\rlap{4} & 20.7 & 1.97   & \\
       \llap{L}MC & \llap{3}eRASS J061532.5-67124\rlap{7} & 13.1 & 0.20 & - \\
       \llap{L}MC & \llap{3}eRASS J062155.1-67010\rlap{8} & 16.4 & 2.10 & - \\
       \llap{L}MC & \llap{1}eRASS J062356.9-67091\rlap{6} &  15.2 & 0.91   & T\\
       \llap{L}MC & \llap{1}eRASS J062435.5-71552\rlap{1} &  18.4 & 0.76  & T\\
       \llap{L}MC & \llap{1}eRASS J063053.8-68490\rlap{4} & 14.4 & 2.90   & T\\
        \hline
    \end{tabular}
    \label{XRstar}
\end{table}

\subsection{AllWISE quality flags}\label{sec:AllWISEflags}

\begin{figure*}
\centering
\begin{tabular}{c}
	\includegraphics[width=\textwidth]{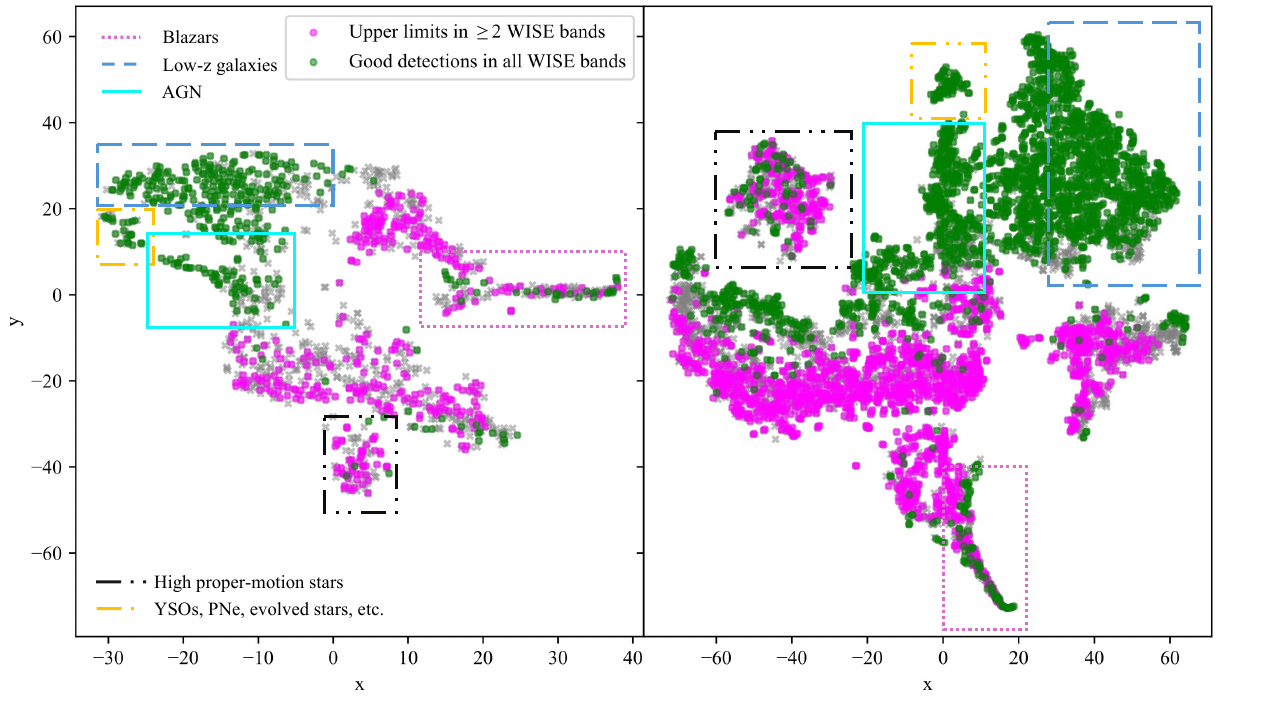}
    \end{tabular}
    \caption{Comparison of the distribution of the quality flags of the photometry from the AllWISE catalogue across the t-SNE maps of the SMC (left) and LMC (right). To be considered a good detection in an individual band the flux signal-to-noise ratio needs to be $>$3. Upper limits are placed on detections with flux signal-to-noise ratio $<$2. The rectangular regions are the same as in Figure \ref{fig:MCtsneclasses}. It can be seen that the majority of sources with "good" detections are in the extragalactic area of the maps.}
    \label{fig:AllWISEflags}
\end{figure*}

The AllWISE catalogue contains quality flags for each of the photometry bands, which are A with a signal-to-noise ratio (SNR)$>$10, B with 3$<$SNR$<$10, C with 2$<$SNR$<$3 and U with SNR$<$2 which is a 95\% confidence upper limit. We kept all sources regardless of the quality flags in order to keep a reasonable sample size of radio detected sources to investigate, as well as to see what the machine learning would pick up on if they were kept in. Figure \ref{fig:AllWISEflags} shows the distribution of the sources with "good" detections (A or B in all bands) and those with "poor" detections (U in 2 or more bands), where he number of "poor" detections is 393/1356 and 2048/6652 for the SMC and LMC, respectively, where most of the upper limits are on the W3 and W4 bands, which do not reach the same depths as the W1 and W2 bands. "Good" detections number 397/1356 and 2679/6652 for the SMC and LMC, respectively. This shows a very distinct divide between "good" and "poor" detections that t-SNE has picked up. Supervised machine learning based classifications from \citet{Pennock2025} show that it is mostly the extragalactic sources that have a "good" detection in all bands and the majority of those deemed as RGB stars or Unknown (see Section \ref{sec:PRF}) are labelled as "poor" and have a maximum of two A or B flag detections out of the full four bands from AllWISE. RGB stars are relatively faint at mid-IR wavelengths, and thus not expected to be detected within AllWISE.

\citet{2019Dalal} showed that various stellar populations can be separated from extragalactic objects in an $J-K_{s}$ vs $K$ CMD diagram. In this, extragalactic objects can be separated from stellar objects approximately using $J-K_{s}>1$ for the LMC and $J-K_{s}>0.9$ for SMC (see region 'L' in figures 2 and 3 in \citet{2019Dalal}). In the 'good' detections sample only 23\% and 11\% of sources in the LMC and SMC, respectively, are predicted to be stellar based of this colour cut, whilst in the 'poor' detections sample 80\% and 61\% of sources are expected to be stellar in origin in the LMC and SMC, respectively. Thus, showing that a significant fraction of the sources in the 'poor' sample are most likely stellar in origin, and unlikely the true counterpart to the radio source.

Restricting the data sample to only those with "good" data would reduce the number of RGB stars that are cross-matched with a radio detection (with the caveat of removing some AGN and galaxies). If the AllWISE "good" detections can be seen as the true radio sources, and the "poor" detections as dominated by Magellanic stars mismatched to radio detections, then AllWISE detections in all bands would make a good prior when matching radio catalogues to optical - IR sources that have better resolution than radio surveys.

\subsection{Other classification methods}\label{sec:otherclass}
Other methods of classification using photometry have been previously employed, and provide useful comparisons to this work. 

\subsubsection{Gaia classifications}
\begin{figure*}
\centering
\begin{tabular}{c}
	\includegraphics[width=\textwidth]{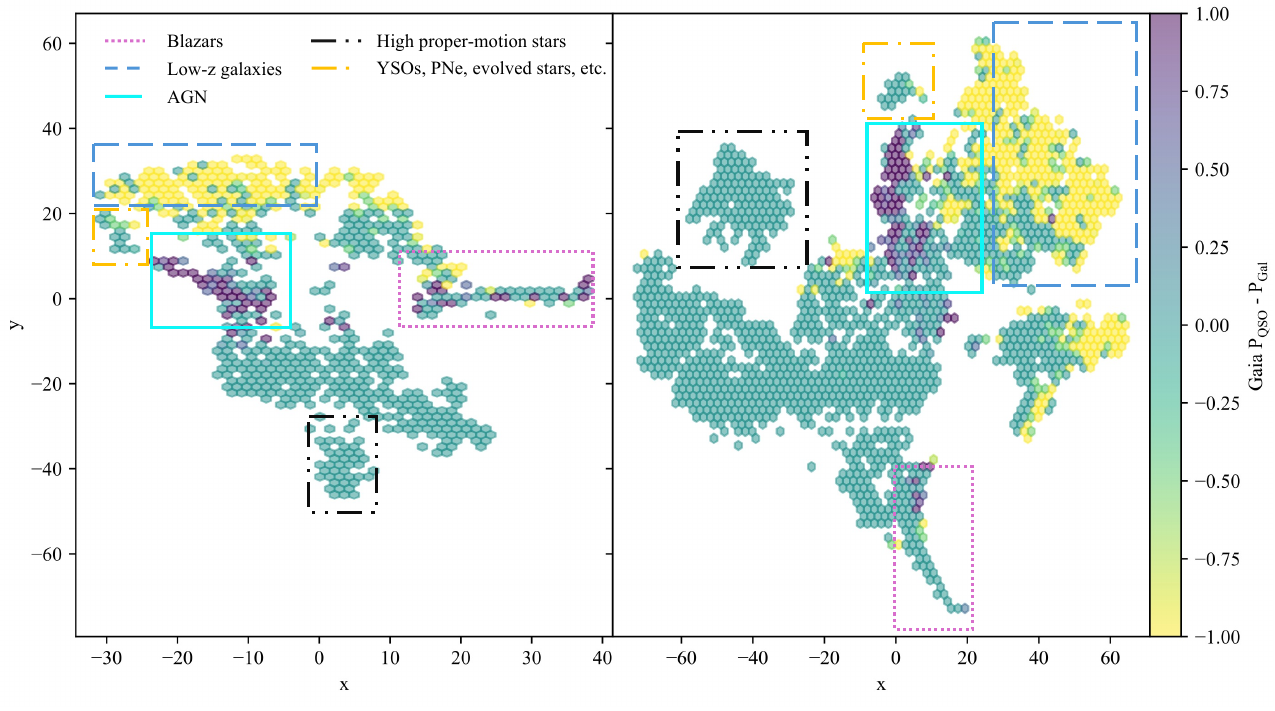}
    \end{tabular}
    \caption{Comparison of the distribution of the median value of Gaia DR3 probabilities of a sources being a QSO, $P_{\rm QSO}$, minus the probability of a source being a galaxy, $P_{\rm Gal}$, in the t-SNE maps of the SMC (left) and LMC (right). $P_{\rm QSO}-P_{\rm Gal}$  roughly leads to: 1 = QSO, 0 = star, $-$1 = galaxy. The dashed rectangular regions are the same as in Figure \ref{fig:MCtsneclasses}.} 
    \label{fig:tsneGaiaQSO}
\end{figure*}

The Gaia DR3 \citep{GaiaDR3} catalogue reports probabilities that the source is a QSO ($P_{\rm QSO}$), a galaxy ($P_{\rm Gal}$) or a stellar source ($P_{\rm Star}$). We investigate how Gaia classification probabilities are distributed across the map, by looking at the distribution of $P_{\rm QSO}-P_{\rm Gal}$, as in Figure \ref{fig:tsneGaiaQSO}. For those sources with the highest probabilities of being a QSO, galaxy or star this leads to the values of $P_{\rm QSO}-P_{\rm Gal}$ close to 1 being highly likely QSOs, those close to $-1$ as highly likely galaxies, and those close to zero as stars. It can be seen that the Gaia QSO classed sources ($P_{\rm QSO}-P_{\rm Gal}\sim1$) match up with most of the known AGN, though do not manage to correctly classify the known AGN that are within the low redshift galaxy region.

Looking at the distribution across the SMC and LMC t-SNE maps of the sources classed as Gaia galaxies ($P_{\rm QSO}-P_{\rm Gal}\sim-1$) it can be seen that this matches up with the areas where known galaxies can be found. Unexpectedly, in the LMC map, there is an area of high probability galaxies in the centre of the t-SNE map (at $x \sim$ $-$13, $y \sim$ 9), to the left of the AGN sequence, which also matches up with high values of $J$ sharp, indicating extended sources (see Figure \ref{tsnemapsVMCsharp}). These sources require further investigation.

The distributions of Gaia classed stars ($P_{\rm QSO}-P_{\rm Gal}\sim0$) across the SMC and LMC t-SNE maps show that stellar sources are found away from where extragalactic sources are found (below $y \sim -$10 for SMC and to the left of $x \sim$ 0 for the LMC; also see Figure \ref{fig:MCtsneclasses}). However, some sources with high probabilities as a stellar source can also be found within the extragalactic areas of the t-SNE maps. For instance, the blazar candidate sequences identified in Figure \ref{fig:MCtsneclasses} are shown to be predicted as mostly stellar sources, though a few are identified as AGN in Figure \ref{fig:tsneGaiaQSO}. 

Comparing with Figure \ref{fig:AllWISEflags} shows that the location of the sources with AllWISE photometry with upper limits aligns with the majority of sources classified in the Gaia catalogue as stars, further corroborating that these sources are most likely not the true radio source.

\subsubsection{Unsupervised vs supervised}\label{sec:PRF}

In \citet{Pennock2025} a supervised machine learning approach was used to classify all the sources in the VMC survey. A probabilistic random forest (PRF) was used, which can handle missing data and therefore, the data did not have to be a "clean" dataset. Also, only the UV to infrared information was used as features, and the radio was left as an independent check after classifying. It was trained on spectroscopically observed sources that were separated into 11 classes: AGN, galaxy, RGB star, OB star, AGB star, post-AGB/RGB star, RSG, PNe, H\,{\sc ii}/YSOs, high-proper-motion/foreground star (PM) and Unknown (Unk).

In Appendix \ref{SALTspec} we present the spectroscopic observations of 12 sources in the direction of the SMC, 6 that were classified by the PRF as a likely (probability$>80\%$) galaxy, and 6 that were classified as a likely AGN. We show that the PRF classifications are correct, finding 6 AGN between the redshifts of 1.08 and 2.34 (Note that we also observed sources that were found in the blazar candidate sequences of the t-SNE maps, where the highest redshift was found to be $z\sim 3.3$, more information on which can be found in Appednix \ref{SALTspec}), and 6 galaxies between the redshifts of 0.037 and 0.089, thus providing evidence of the accuracy of the PRF classifier. 

\begin{figure*}
\centering
\begin{tabular}{c}
	\includegraphics[width=\textwidth]{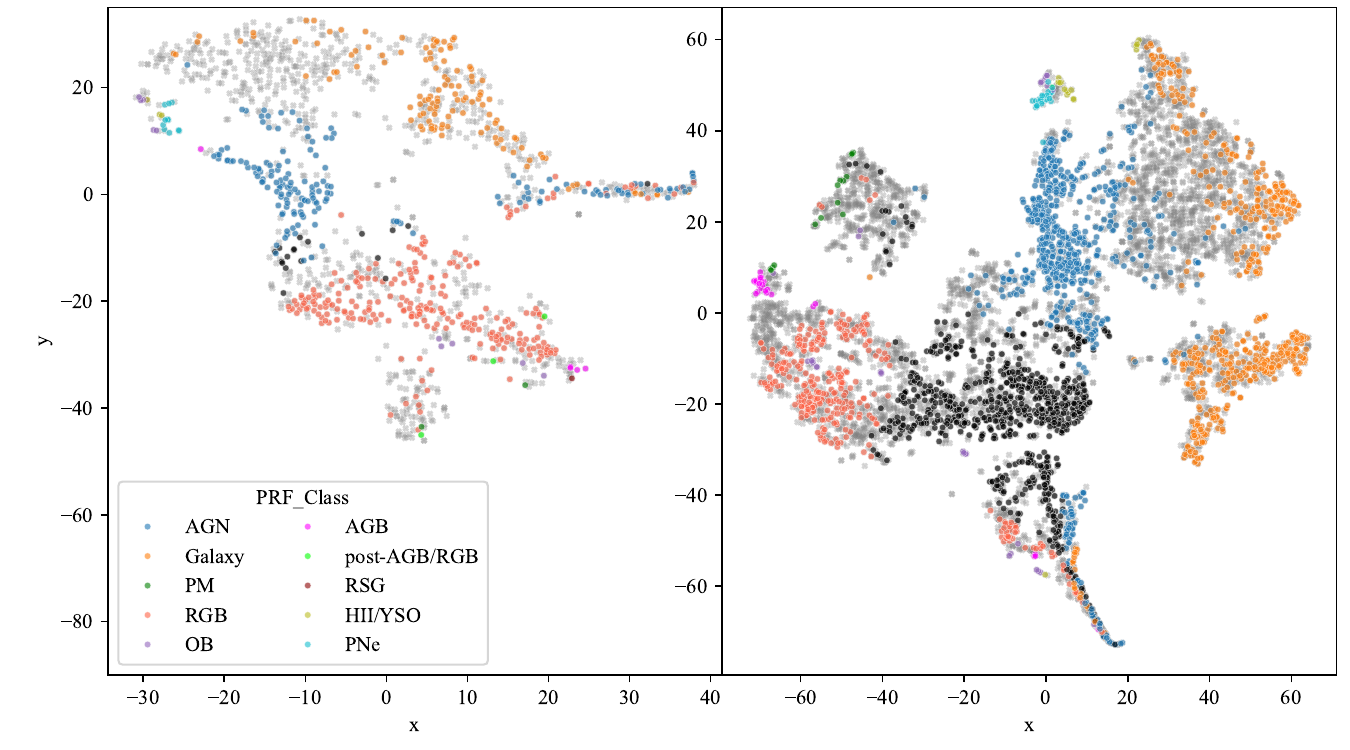}
    \end{tabular}
    \caption{Classifications of the t-SNE map sources of the SMC (left) and LMC (right) made by a PRF classifier. Grey crosses represent sources without a high probability ($>$ 80\%) for a particular class and black dots represent sources that have been classified as "Unknown". The PRF classifier separated sources into 11 classes: AGN, galaxy, RGB (red-giant branch) star, OB (type-O and type-B) star, AGB (asymptotic giant branch) star, post-AGB/RGB star, RSG (red supergiant), PNe (planetary nebul\ae\ ), H\,{\sc ii}/YSOs (young stellar objects), high-proper-motion/foreground star (PM) and Unknown.}
    \label{fig:tsnePRF}
\end{figure*}

Testing the PRF classifiers produced in that work on the radio population of the data shows, in Figure \ref{fig:tsnePRF}, that the classes are grouped where expected on the t-SNE maps for those with a probability of classification $> 80$\%. As before, the AGN are concentrated across the middle of the t-SNE maps (SMC:y$\sim$-10 -- 10; LMC: x$\sim$0 -- 20) and, as expected, some can be found towards the galaxy population (towards the top in the SMC and right in the LMC), with galaxies becoming more dominant the further away from the AGN population. 

RGB stars are not expected to be radio sources, and so these are likely cases of a superposition of a star in front of the true radio source, or the star being the closest cross-match available as the true radio source is undetected at near-IR wavelengths. 

"Unknown" classifications (black points) appear to be placed between AGN classifications and RGB classifications, as well as in the separate island of high proper-motion sources. These could be cases of a wrong match like for the RGB classification, or a sub-class of the extragalactic that has not been accounted for in the training of the PRF classifiers. Another possibility is that these are variable objects, which would mean that, as the various surveys used in this work were not observed at the same time, the SED is not a true reflection of the underlying source, and would not match the 'expected' of any classification that is of a less variable object. However, comparing with Figure \ref{fig:VMCVAR} shows that whilst varying sources can be found here, they generally have low variability probabilities.
Of the eleven that had a Simbad entry, five were "Stars" and the rest were known variable stars, and in one case a BL Lacertae object -- a type of AGN. 

Comparing Figure \ref{fig:tsnePRF} with Figure \ref{fig:AllWISEflags} shows that there is a population of galaxies (orange points) that match up with the sources with upper limits on $\geq$ 2 WISE bands, that can be found between the low-$z$ galaxies and blazar populations. Only 49/667 of the sources for this cluster in the LMC have good detections in all WISE bands, and 284/667 have upper limits in $\geq$ 2 WISE bands. Galaxies that are faint in mid-IR could be indications of a lack of dust or star formation.

Overall, the comparison between the unsupervised t-SNE and supervised PRF shows that the two methods separate the different types of sources well. The PRF also provides insight into the possible classifications of the sources in the areas of the t-SNE maps that we do not have spectroscopic classifications for. 

%%%%%%%%%%%%%%%%%%%%%%%%%%%%%%%%%%%%%%%%%%%%%%%%%%%%%%%%%%%%
\subsection{The blazar candidate sequence spectroscopic follow-up}\label{blazseq}

We selected nine sources in the direction of the LMC and nine sources in the direction of the SMC that were found in the blazar candidate sequence of the t-SNE maps (see Figure \ref{fig:MCtsneclasses}) and bright enough to be observed in the optical with SALT (see Appendix \ref{SALTspec} for more information and Figure \ref{fig:tsnePRF_blaz} for their locations on the t-SNE maps). From the LMC sample we found five extragalactic sources, two Magellanic sources and two unconfirmed sources due to poor signal-to-noise. From the SMC sample we found three extragalactic sources, four Magellanic sources and two unconfirmed due to poor signal-to-noise. It is not surprising some stellar sources (most likely not the true counterparts to the radio sources) made it into the sample as from Figure \ref{tsnemapsPM} we can see that some sources with PM$>$4 milli-arcsec/yr can be found amongst the sources with little to no proper-motions. 

Figure \ref{fig:tsnePRF} shows that sources classified as 'RGB' stars by the PRF can be found amongst the blazar candidate sequence. The PRF classifies the SMC sources (with probabilities $\sim$0.22--0.38) SMC\,tSNE\,b1 , b2, b3 and b4 as AGN, SMC\,tSNE\,b5, b6 and b7 as galaxies, SMC\,tSNE\,b8 as an RGB star and SMC\,tSNE\,b9 as an OB star. The PRF classifies the LMC source (with probabilities $\sim$0.58--0.97) LMC\,tSNE\,b1 as an H\,\textsc{ii}/YSO, which matches with it being an emission line object, LMC\,tSNE\,b2 and b7 as galaxies and LMC\,tSNE\,b3, b4, b5, b8 and b9 as AGN. Lastly, LMC\,tSNE\,b6 is classified as Unknown, which is expected since its spectrum reveals it is an early-M dwarf, which the classifier was not trained on.

Figure \ref{fig:AllWISEflags} shows that sources with poor detections in AllWISE are also amongst this sample, which tend to be an indication of sources of stellar origin (see Section \ref{sec:AllWISEflags}). Sources with "good" detections in AllWISE are SMC\,tSNE\,b1, b2, b3 and LMC\,tSNE\,b1, b2, b3, b4, b5 and b9, which are all AGN or galaxies, with the exception of LMC\,tSNE\,b1 which is an emission line object and LMC\,tSNE\,b9 which is unknown, but good detections in AllWISE would indicate an extragalactic object is more likely. The "poor" detections are SMC\,tSNE\,b4, b7, an unknown and an RGB star, and LMC\,tSNE\,b6, an early M-dwarf star whose spectrum would not be expected to be bright in the reddest AllWISE bands.

The primary spectral signature to look for in a blazar candidate \citep[see, e.g.][]{2026Nlowie} is a steep blue non-thermal continuum indicating optical jet dominance, often with weak AGN or diluted host galaxy features. This is not apparent in most of these spectra as many show strong broad emission lines with relatively normal QSO-like continua, which would place them closer to radio-quiet AGN or standard flat-spectrum radio quasars (FSRQs), a class of blazars without a dominant jet component. The two exceptions to this are SMC\,tSNE\,b3 ($z$ = 1.643) and LMC\,tSNE\,b4 ($z$ = 0.522), which show steep blue continuum combined with strong emission features and minimal host galaxy contribution, broadly consistent with a jet-dominated FSRQ.

Blazars are expected to be highly variable on timescales as short as hours even in the optical due to fast fluctuations in the jet \citep[e.g.][]{2025Raiteri}. From looking at the light-curves in the VMC K$_s$ band (see Figure \ref{SMCvar} and \ref{LMCvar}) of the extragalactic and unconfirmed sources we find that LMC\,tSNE\,b5 ($z$=1.50) and  LMC\,tSNE\,b8 ($z$=2.80) clearly show the slow variation in brightness commensurate with an AGN \citep[e.g.][]{padovani2017}. The rest of the sources, bar one, show little to no variation. The exception is LMC\,tSNE\,9 (unconfirmed class/redshift), which appears to show rapid variability between $K_\mathrm{s}\sim15$--$16$ mag, which could be an indication that this source is actually a blazar.

Gamma-ray detections are also a good indication of a blazar, as they are known to dominate the extragalactic $\gamma$-ray sky \citep{1995PASP..107..803U,2009ApJ...697.1071A,2015ApJ...810...14A}. To see if any of these sources were $\gamma$-ray detected we cross-matched the blazar sample with the 4FGL-DR4 catalogue (see \citet{2009ApJ...697.1071A, 2026ApJ...999..244N} and Section 2.1 of \citet{2026Nlowie}) with a 30$^{\prime\prime}$ search area (\textit{Fermi}-LAT has arcminute positional uncertainties and typically $\sigma_{pos}\sim30^{\prime\prime}$ can be used for the 95\% confidence limit; \citet{2023Ballet}), from which we obtain four matches. SMC\,tSNE\,b2 ($z$=1.812) has the furthest separation from a $\gamma$-ray detection, and its optical spectrum shows clear strong broad-line features, typical of a type 1 AGN, so this is perhaps not a blazar or even the true counterpart to the $\gamma$-ray detection. SMC\,tSNE\,b9 is clearly an A-type star from its optical spectrum, so this is most likely not the true counterpart either. LMC\,tSNE\,b4 ($z$ = 0.522) shows the expected steep blue continuum in the optical spectrum of a blazar, so is more likely the correct counterpart to the $\gamma$-ray detection (4FGL\,J0509.9$-$6417), where the 4FGL catalogue provides a classification of a BL Lac. Lastly, the closest cross-match of $0\rlap{.}^{\prime\prime}53$ is of 4FGL\,J0443.3$-$6652 with LMC\,tSNE\,b9, a blazar which is classified as an FSRQ. Combining this with its highly variable nature, good detections in AllWISE and the PRF classification as an AGN adds to the certainty of this classification. Further to this, \citet{2026A&A...709A..37H} created a catalogue of blazar and blazar candidates from the literature and matched it with the eRASS1 catalogue. From the 18 sources, LMC\,tSNE\,b4 and LMC\,tSNE\,b9 are within this blazar catalogue, identified as FSRQs with redshifts $z=0.518$ and $z=0.746$, respectively.

\begin{table}
    \centering
    \caption{Results of the cross-match between the 18 spectroscopically observed blazar candidate sample (see Appendix \ref{SALTspec}, for more information) and the 4FGL catalogue with a 30$^{\prime\prime}$ search area, providing the separation, IDs and classification of the sources in the 4FGL catalogue.}
    \begin{tabular}{cccc}
    \hline \hline
        ID & separation ($^{\prime\prime}$) & 4FGL Name & 4FGL class\\
        \hline
        SMC\,tSNE\,b2 & 27.5 & 4FGL\,J0046.7$-$7048 & Unknown \\
        SMC\,tSNE\,b9 & 17.5 & 4FGL\,J0058.0$-$7245e & Galaxy \\
        LMC\,tSNE\,b4 & 19.1 & 4FGL\,J0509.9$-$6417 & BL Lac\\
        LMC\,tSNE\,b9 & 0.5 & 4FGL\,J0443.3$-$6652 & FSRQ\\
        \hline
    \end{tabular}

    \label{tab:gamma-blazar}
\end{table}

Overall, the blazar candidate sequence of the t-SNE maps are not purely made up of blazars, but with other information such as variability, $\gamma$-ray detection, measured proper motions, PRF classifications and reliability flags of the AllWISE catalogue we can narrow down the sample to the most likely blazar candidates.

%%%%%%%%%%%%%%%%%%%%%%%%%%%%%%%%%%%%%%%%%%%%%%%%%%%%%%%%%%%%
%%%%%%%%%%%%%%%%%%%%%%%%%%%%%%%%%%%%%%%%%%%%%%%%%%%%%%%%%%%%

\section{Summary and conclusions}\label{sec:conc}
We used a t-SNE algorithm on multi-wavelength data from optical to radio of the 8,008 radio detected sources in the Magellanic Clouds. This reduced the dimensions to two, whilst maintaining the local structure of the data in order to cluster similar objects together in 2D space.

After applying the t-SNE algorithm we find:
\begin{itemize}
    \item Extragalactic, Magellanic and foreground objects separate from each other into noticeable clusters, with some overlap. Foreground stars tend to be clustered away from the rest of the sources. Dusty evolved Magellanic stars also form their own separate cluster, though there is some overlap with the nearby AGN cluster. AGN tend to be found across the centres of the map, at one end of which a population of blazars can be found. Some AGN encroach on the galaxies, which is not unexpected. Other Magellanic stellar sources are clustered next to and away from the the AGN sequence.
    \item The Gaia BP/RP excess and VISTA sharpness values trace how extended an object is, causing a gradient in redshift across the extragalactic population on the t-SNE map.
    \item An increased probability of a source being variable in the near-IR can be found in the region of the t-SNE with known AGN, which are expected to be variable sources. Increased variability also aligns with the more extended galaxies, which could indicate a low luminosity AGN, or that the extended profile is causing errors in the photometry used to calculate the variability.
    \item The majority of X-ray detected sources are found where AGN are predicted to lie, as expected, including a blazar candidate sequence.
    \item AllWISE detections that were upper limits were not removed when collating the photometry. Sources that have upper limits in $\geq$ 2 WISE bands are clearly separated from those that have good detections in all bands. The good detections align with the majority of the extragalactic sources, as well as the dusty stellar sources (e.g. YSOs, PNe) that emit in the mid-IR. The poor detections align with foreground stars, Magellanic stars and sources that are unknown, which could imply that these are not the true radio sources, and that requiring good detections in WISE bands is a good metric for determining if an optical/IR source is the true match to the radio detection.
    \item Previous classifications from Gaia and supervised machine learning \citep{Pennock2025} align well with the clusters in the t-SNE maps. Combining the supervised with the unsupervised methods helps to inform the possible identities of some of the clusters for which there are no spectroscopic classifications available. 
    \item Follow-up optical spectroscopy of the blazar candidate sequence with SALT has revealed some strong blazar candidates can be found with the t-SNE map. However, the presence of stars and type 1 AGN show that further information is needed alongside the t-SNE to choose the more likely blazar candidates.
\end{itemize}

We have thus demonstrated broad agreement between different classification schemes -- t-SNE, PRF, Gaia and spectroscopic -- which can complement one another to obtain more complete samples and be used in conjunction to obtain purer samples.

Future work could involve removing sources with only upper limits in WISE data and seeing how this affects the clustering. On the other hand, it is possible that these upper limits helped separate the extragalactic from the stellar, so another avenue of exploration would be to put back in the sources that had missing data from any band and provide an upper limit for the missing data, which would then increase the number of radio detected sources we are classifying in the direction of the Magellanic Clouds. Another avenue of exploration would be to utilise this technique on the stellar dense field of the Milky Way.

\section*{Acknowledgements}
We thank the anonymous referee for their feedback, which helped improve the paper. C.M.P. acknowledges funding from a UKRI Future Leaders Fellowship (grant codes MR/T020989/1 and MR/Y019539/1). M.I.N. acknowledges support from the Development in Africa with Radio Astronomy (DARA) project through an MSc by Research studentship. H.A. benefitted from grant CIIC 165/2026 of DAIP, Univ.\ de Guanajuato. This research was funded in part by the National Science Centre (NCN), Poland (grant number 2023/49/B/ST9/00066). 

Some of the observations reported in this paper were obtained with the Southern African Large Telescope (SALT).

For the purpose of open access, the author has applied a Creative Commons Attribution (CC BY) licence to any Author Accepted Manuscript version arising from this submission.

NOIRLab IRAF is distributed by the Community Science and Data Center at NSF NOIRLab, which is managed by the Association of Universities for Research in Astronomy (AURA) under a cooperative agreement with the U.S. National Science Foundation.

This research has made use of the publicly available programming language P\textsc{ython}, including N\textsc{um}P\textsc{y} \citep{Numpy}, S\textsc{ci}P\textsc{y} \citep{Scipy}, M\textsc{atplotlib} \citep{Matplotlib}, A\textsc{stropy} \citep{2013Astropy}, S\textsc{cikit-learn} \citep{scikit-learn} and the T\textsc{opcat} analysis program \citep{Taylor2011}.

%%%%%%%%%%%%%%%%%%%%%%%%%%%%%%%%%%%%%%%%%%%%%%%%%%
\section*{Data Availability}

%The inclusion of a Data Availability Statement is a requirement for articles published in MNRAS. Data Availability Statements provide a standardised format for readers to understand the availability of data underlying the research results described in the article. The statement may refer to original data generated in the course of the study or to third-party data analysed in the article. The statement should describe and provide means of access, where possible, by linking to the data or providing the required accession numbers for the relevant databases or DOIs.

The data underlying this article will be shared on reasonable request to the corresponding author.

%%%%%%%%%%%%%%%%%%%% REFERENCES %%%%%%%%%%%%%%%%%%

% The best way to enter references is to use BibTeX:

\bibliographystyle{mnras}
\bibliography{example} % if your bibtex file is called example.bib

% Alternatively you could enter them by hand, like this:
% This method is tedious and prone to error if you have lots of references
%\begin{thebibliography}{99}
%\bibitem[\protect\citeauthoryear{Author}{2012}]{Author2012}
%Author A.~N., 2013, Journal of Improbable Astronomy, 1, 1
%\bibitem[\protect\citeauthoryear{Others}{2013}]{Others2013}
%Others S., 2012, Journal of Interesting Stuff, 17, 198
%\end{thebibliography}

%%%%%%%%%%%%%%%%%%%%%%%%%%%%%%%%%%%%%%%%%%%%%%%%%%

%%%%%%%%%%%%%%%%% APPENDICES %%%%%%%%%%%%%%%%%%%%%

\appendix

\section{SALT spectroscopy}\label{SALTspec}

Table~\ref{table:SALTobs} lists the long-slit spectroscopic observations obtained with the Robert Stobie Spectrograph \citep[RSS:][]{2003SPIE.4841.1463B,2003SPIE.4841.1634K} on the Southern African Large Telescope \citep[SALT:][]{2006SPIE.6267E..0ZB} at Sutherland, South Africa, in priority-4 (filler) programmes 2023-1-SCI-018, 2023-2-SCI-014, 2024-1-SCI-027, 2024-2-SCI-029, 2025-1-SCI-012 and 2025-2-SCI-018 (PI for all: Jacco van Loon). Three sets of targets were selected: from the suspected blazar sequence on the tSNE maps of the SMC and LMC fields, respectively, and AGN and galaxy candidates classified as such by the PRF (in the SMC field). They were generally 17--18 mag in the optical (the faintest was LMC\,tSNE\,b9 with $G$ = 18.6 mag), with a few brighter sources. Multiple spectra of the same target were averaged except where noted because of poor quality. The slit position angle (measured from North over East) was kept at 0 (North--South aligned) except where noted otherwise so as to cross another object of (potential) interest.

\begin{table*}
\caption{Log of SALT spectroscopy.}
\vspace{5pt}
\begin{tabular}{llllllllll}
\hline\hline
Target & RA (J2000) & Dec (J2000) & Programme ID & Date & Airmass & Seeing & Cloud & Moon & Remarks \\
& $\mathrm{h\ m\ s}$ & $\circ\ \prime\ \prime\prime$ & & dd/mm/yy & & $\prime\prime$ & & phase & \\
\hline
\multicolumn{10}{l}{\em SMC tSNE blazar sequence} \\
SMC\,tSNE\,b1 & 0 56 36.77 & $-$74 03 15.2 & 2023-1-SCI-018 & 05/10/23 & 1.380 & $>$ 3 & thin & 0.56 & \\
SMC\,tSNE\,b2 & 0 41 25.35 & $-$70 57 43.1 & 2023-1-SCI-018 & 29/08/23 & 1.332 & 2.1 & clear & 0.98 & (not used) \\
& & & 2024-1-SCI-027 & 02/10/24 & 1.317 & 4.0 & thin & 0.00 & \\
SMC\,tSNE\,b3 & 1 03 31.77 & $-$70 06 38.0 & 2023-1-SCI-018 & 05/10/23 & 1.282 & $>$ 3 & thin & 0.56 & \\
SMC\,tSNE\,b4 & 0 56 33.06 & $-$73 57 10.1 & 2023-1-SCI-018 & 29/08/23 & 1.337 & 1.9 & clear & 0.98 & low signal \\
SMC\,tSNE\,b5 & 0 44 14.35 & $-$72 43 07.2 & 2023-1-SCI-018 & 05/10/23 & 1.323 & $>$ 3 & thin & 0.56 & 600 s\\
SMC\,tSNE\,b6 & 0 51 41.26 & $-$72 56 00.7 & 2023-1-SCI-018 & 14/07/23 & 1.316 & 2.3 & thin & 0.07 & 300 s \\
SMC\,tSNE\,b7 & 1 10 39.29 & $-$72 28 24.4 & 2023-1-SCI-018 & 14/07/23 & 1.316 & 2.3 & thin & 0.07 & 400 s, PA = $-$30$^\circ$ \\
SMC\,tSNE\,b8 & 0 46 03.81 & $-$74 13 29.9 & 2023-1-SCI-018 & 28/08/23 & 1.350 & 2.3 & clear & 0.93 & \\
SMC\,tSNE\,b9 & 0 55 32.83 & $-$72 31 23.8 & 2023-1-SCI-018 & 14/07/23 & 1.307 & 2.6 & thin & 0.07 & 600 s, PA = $+$90$^\circ$ \\
\hline
\multicolumn{10}{l}{\em LMC tSNE blazar sequence} \\
LMC\,tSNE\,b1 & 4 57 16.23 & $-$66 23 20.8 & 2024-2-SCI-029 & 06/11/24 & 1.271 & 2.7 & thin & 0.27 & PA = $+$85$^\circ$ \\
LMC\,tSNE\,b2 & 6 04 43.70 & $-$65 16 47.1 & 2024-2-SCI-029 & 19/11/24 & 1.194 & 3.5 & clear & 0.87 & PA = $-$14$\rlap{.}^\circ$5 \\
LMC\,tSNE\,b3 & 5 54 35.86 & $-$71 01 54.3 & 2024-2-SCI-029 & 06/11/24 & 1.280 & 3.5 & thin & 0.27 & \\
LMC\,tSNE\,b4 & 5 08 20.39 & $-$64 33 17.5 & 2023-2-SCI-014 & 28/11/23 & 1.189 & 2.0 & thin & 0.97 & \\
LMC\,tSNE\,b5 & 5 18 03.02 & $-$63 25 59.4 & 2024-2-SCI-029 & 19/11/24 & 1.177 & 3.5 & clear & 0.87 & \\
LMC\,tSNE\,b6 & 4 49 37.52 & $-$72 16 29.5 & 2024-2-SCI-029 & 19/11/24 & 1.358 & 2.5 & clear & 0.87 & PA = $-$2$^\circ$ \\
LMC\,tSNE\,b7 & 6 11 52.56 & $-$66 38 02.1 & 2024-2-SCI-029 & 15/01/25 & 1.266 & 3.6 & clear & 0.95 & (not used) \\
 & & & 2025-2-SCI-018 & 08/12/25 & 1.220 & 1.1 & clear & 0.79 & \\
 & & & 2025-2-SCI-018 & 10/12/25 & 1.210 & 1.5 & clear & 0.59 & (not used) \\
 & & & 2025-2-SCI-018 & 18/12/25 & 1.215 & 1.4 & clear & 0.01 & \\
 & & & 2025-2-SCI-018 & 21/12/25 & 1.284 & 1.8 & clear & 0.04 & \\
LMC\,tSNE\,b8 & 6 22 41.86 & $-$68 59 21.9 & 2023-2-SCI-014 & 22/01/24 & 1.248 & 3.3 & clear & 0.93 & (not used) \\
 & & & 2024-2-SCI-029 & 02/01/25 & 1.305 & 3.3 & thin & 0.11 & \\
 & & & 2024-2-SCI-029 & 02/01/25 & 1.368 & 2.5 & thin & 0.11 & \\
LMC\,tSNE\,b9 & 4 43 18.24 & $-$66 52 04.5 & 2023-2-SCI-014 & 03/01/24 & 1.310 & 2.5 & clear & 0.52 & (not used) 750 s \\
& & & 2024-2-SCI-029 & 07/11/24 & 1.227 & 1.4 & thick & 0.37 & \\
& & & 2024-2-SCI-029 & 15/02/25 & 1.341 & 2.0 & clear & 0.89 & \\
& & & 2025-1-SCI-012 & 09/06/25 & 1.305 & 1.7 & clear & 0.99 & \\
\hline
\multicolumn{10}{l}{\em SMC PRF follow up ("a" = likely AGN; "g" = likely galaxy)} \\
SMC\,PRF\,a1 & 1 35 05.07 & $-$75 07 03.4 & 2024-1-SCI-027 & 02/10/24 & 1.362 & 2.6 & thin & 0.00 & \\
SMC\,PRF\,a2 & 0 38 32.36 & $-$70 48 46.8 & 2024-2-SCI-029 & 11/11/24 & 1.348 & 2.3 & clear & 0.80 & \\
SMC\,PRF\,a3 & 0 19 03.85 & $-$74 29 42.9 & 2025-1-SCI-012 & 02/10/25 & 1.363 & 3.5 & clear & 0.79 & \\
SMC\,PRF\,a4 & 0 40 32.76 & $-$75 24 27.3 & 2025-1-SCI-012 & 02/10/25 & 1.367 & 3.3 & clear & 0.79 & \\
SMC\,PRF\,a5 & 0 27 00.95 & $-$73 57 42.9 & 2024-1-SCI-027 & 02/10/24 & 1.336 & 3.5 & thin & 0.00 & \\
SMC\,PRF\,a6 & 1 03 31.72 & $-$70 06 37.0 & 2025-1-SCI-012 & 09/07/25 & 1.346 & 1.7 & clear & 0.99 & \\
SMC\,PRF\,g1 & 0 55 49.13 & $-$74 30 58.3 & 2025-1-SCI-012 & 09/07/25 & 1.355 & 1.9 & clear & 0.99 & PA = $+$70$^\circ$ \\
SMC\,PRF\,g2 & 1 35 45.78 & $-$72 13 13.7 & 2025-1-SCI-012 & 09/07/25 & 1.317 & 1.6 & clear & 0.99 & PA = $-$27$^\circ$ \\
SMC\,PRF\,g3 & 1 45 15.60 & $-$71 02 08.1 & 2024-1-SCI-027 & 10/10/24 & 1.283 & 1.6 & thick & 0.52 & \\
SMC\,PRF\,g4 & 1 41 12.32 & $-$72 03 09.3 & 2025-1-SCI-012 & 02/10/25 & 1.298 & 3.7 & clear & 0.79 & \\
SMC\,PRF\,g5 & 0 51 16.89 & $-$73 40 01.4 & 2025-1-SCI-012 & 02/10/25 & 1.350 & 3.5 & clear & 0.79 & PA = $-$45$^\circ$ \\
SMC\,PRF\,g6 & 0 57 45.52 & $-$74 27 11.7 & 2025-1-SCI-012 & 02/10/25 & 1.360 & 3.5 & clear & 0.79 & \\
\hline
\end{tabular}
\label{table:SALTobs}
\end{table*}

We used the low resolution grating PG0700, with order blocking filter PC03400, and a 1$\rlap{.}^{\prime\prime}$25 wide (and 8$^\prime$ long) slit, 2$\times$2 on-chip binning (resulting in a pixel scale of 1.23 \AA\ in the dispersion direction and 0$\rlap{.}^{\prime\prime}$253 in the spatial direction), FAINT gain and SLOW readout, with wavelength calibration based on 2-s exposure of an argon arc lamp -- except in programmes 2023-1-SCI-018 and 2023-2-SCI-014 when a 1$\rlap{.}^{\prime\prime}$5 slit was used and 7.5-s exposures of a copper--argon arc lamp. Exposure times were 900 s except where noted otherwise.

Data were received after having been processed with the PySALT pipeline\footnote{http://pysalt.salt.ac.za/} \citep{2010SPIE.7737E..25C}: corrections for gain, cross-talk, bias; mosaicked, replacing bad pixels, cleaning cosmic rays, filling CCD gap and calibrating the wavelength axis. We then used standard routines in IRAF \citep{1986SPIE..627..733T,1993ASPC...52..173T,2025ASPC..541..461F} to subtract the sky background after fitting a second-order Chebyshev function across two bands on the CCD at either side of the target, and then extract the spectrum using a variance-weighted spatial-profile and fitting a second-, third- or fourth-order Legendre polynomial to trace the spectrum along the dispersion. The response was partially compensated by using a response function derived uing a similar set up \citep{2026MNRAS.546ag012V}.

\begin{figure*}
\centering
\begin{tabular}{c}
	\includegraphics[width=0.78\textwidth]{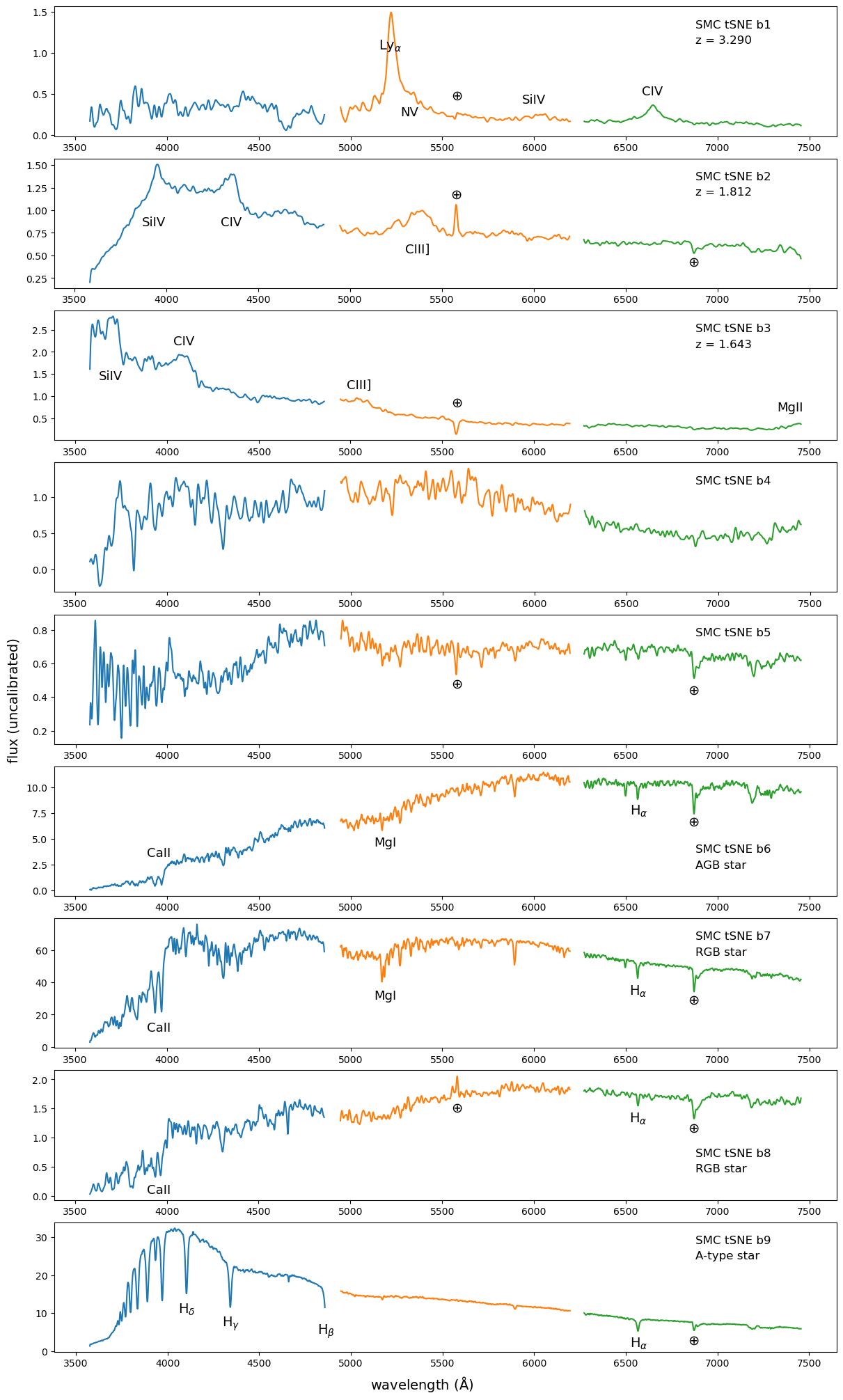}
    \end{tabular}
    \caption{Spectra of sources on the SMC tSNE blazar candidate sequence.}
    \label{fig:SMCtSNEb}
\end{figure*}

\begin{figure*}
\centering
\begin{tabular}{c}
	\includegraphics[width=0.78\textwidth]{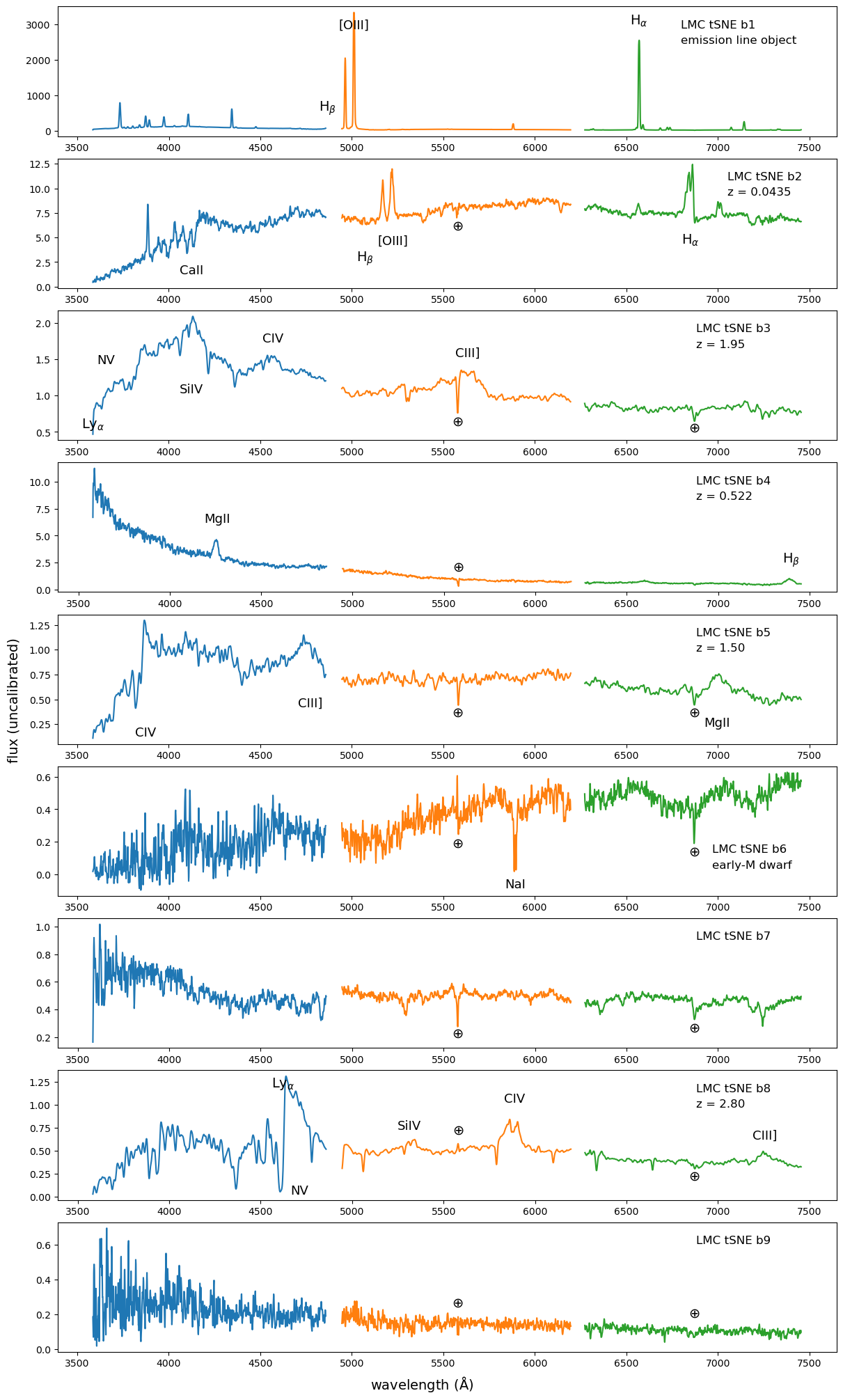}
    \end{tabular}
    \caption{Spectra of sources on the LMC tSNE blazar candidate sequence.}
    \label{fig:LMCtSNEb}
\end{figure*}

\begin{figure*}
\centering
\begin{tabular}{c}
	\includegraphics[width=0.93\textwidth]{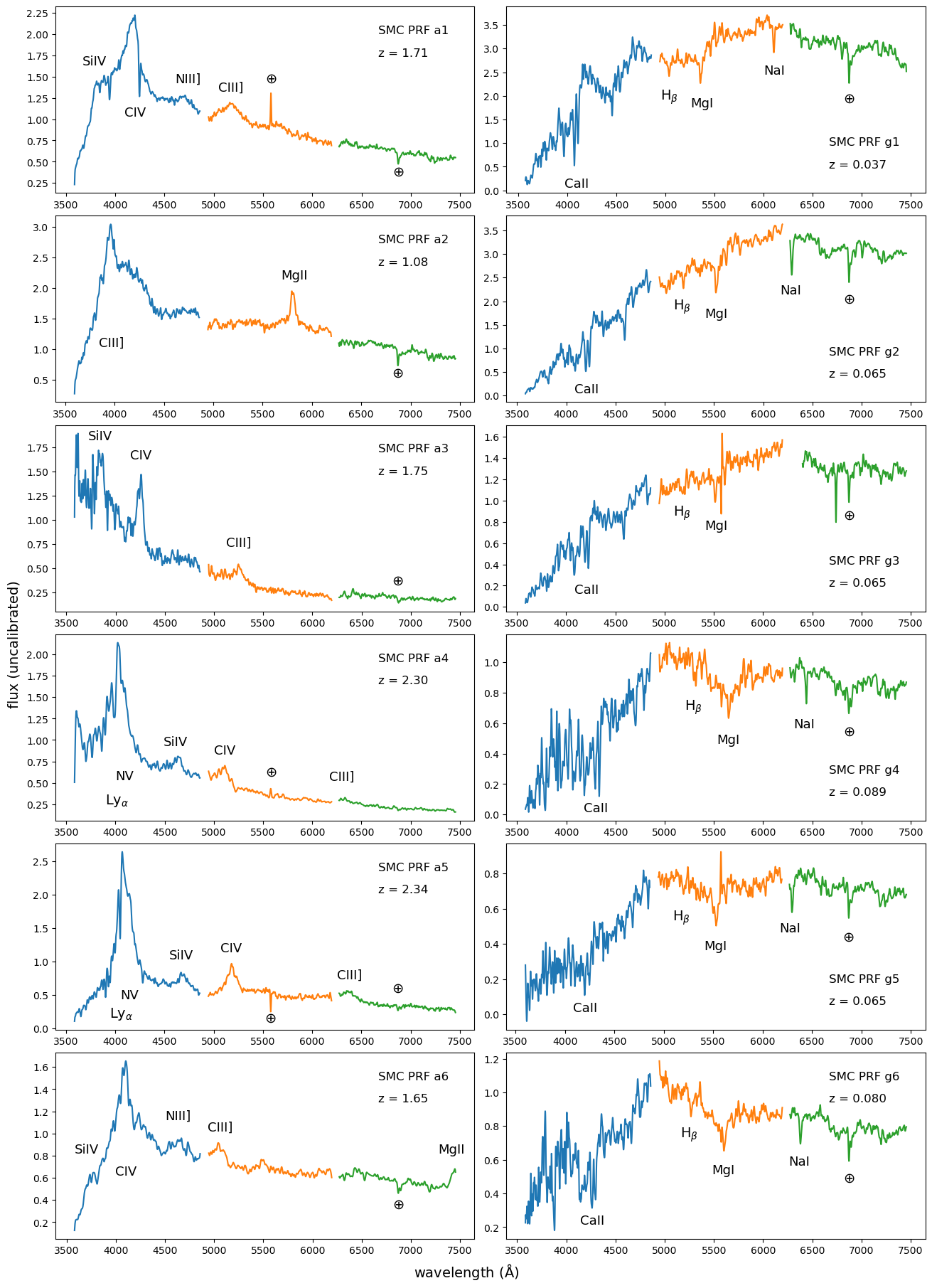}
    \end{tabular}
    \caption{Spectra of sources classified by the PRF as `AGN' or `Galaxy'.}
    \label{fig:SMCprf}
\end{figure*}

The results are displayed in figures~\ref{fig:SMCtSNEb}, \ref{fig:LMCtSNEb} and \ref{fig:SMCprf} and summarised in table~\ref{table:SALTclass}; note that SMC\,tSNE\,b3 and SMC\,PRF\,a6 are the same source (but independent spectra), by sheer coincidence. In several t-SNE cases the radio source was misidentified with a star; the true t-SNE source is likely a radio galaxy or AGN blended by a foreground star. In the case of LMC\,tSNE\,b1 the radio emission likely arises in the ionised gas of the N\,11 H\,{\sc ii} region in the LMC (the H$\alpha$, H$\beta$ and [O\,{\sc iii}] lines saturated in the core); in the case of SMC\,tSNE\,b9 the hitherto anonymous A-type star (Gaia DR3 4688992925848942080) has a red giant companion \citep[blended source SSTISAGEMA J005532.19$-$723123.1;][]{2011AJ....142..103B}. \citet{2021AJ....162..177P} present the optical spectrum of LMC\,tSNE\,b9, which shows one narrow line which they attribute to Mg\,{\sc i} 2798\AA. Unfortunately it ended up precisely in the CCD gap of our SALT spectrum, at 4900 \AA, but the hot, featureless continuum indeed suggests a blazar-like AGN. Overall, the PRF classifications of this sample are found to be 100 per cent correct, both for 6 AGN and 6 galaxies. Five AGN found by Gaia were not known at the time of observation preparation, but we confirm all of them and find several new AGN not found by Gaia, in the 1.1--3.3 redshift range, which is higher than the t-SNE selected sample studied in \citet{2022MNRAS.515.6046P}. We also find hitherto unknown low-redshift galaxies, as near as $z=0.037$. Note that most of the PRF sources are very faint/undetected in radio, except for SMC\,PRF\,g5, which is found to be a $\sim$6.3$^\prime$ long WAT (wide-angle tailed) radio galaxy in MeerKAT images \citep[see][]{2024Cotton},  which also coincides with an X-ray detection, which could possibly be due to a cluster.

\begin{table*}
\caption{Source classifications.}
\vspace{5pt}
\begin{tabular}{lllll}
\hline\hline
Target & Known source & \multicolumn{3}{c}{\rule{35mm}{0.2mm}\ Classification\ \rule{35mm}{0.2mm}} \\
 & & Literature & ref. & This work \\
 \hline
SMC\,tSNE\,b1 & [FBR2002] J005636$-$740316 & Radio source & 1 & AGN $z=3.290$ \\
SMC\,tSNE\,b2 & Gaia DR3 4690076391421961472 & AGN $z=1.808$ & 2 & AGN $z=1.812$ \\
SMC\,tSNE\,b3 & PMN J0103$-$7006 & & 3 & AGN $z=1.643$ \\
SMC\,tSNE\,b4 & 2MASS J00563305$-$7357109 & RGB star & 4 & ? \\
SMC\,tSNE\,b5 & 2MASS J00441430$-$7243076 & RGB star & 4,5 & ? \\
SMC\,tSNE\,b6 & 2MASS J00514127$-$7256003 & Red supergiant & 4 & AGB star (not radio?) \\
SMC\,tSNE\,b7 & XMMU J011040.9$-$722830 & Radio galaxy & 6 & RGB star (unrelated?) \\
SMC\,tSNE\,b8 & 2MASS J00460387$-$7413294 & RGB star & 4 & RGB star (not radio?) \\
SMC\,tSNE\,b9 & [FBR2002] J005531$-$723122 & Radio source & 1 & A-type star (unrelated?) \\
LMC\,tSNE\,b1 & [ELS2006] N11 028, HD 32340 & Mid-O-type star & 7,8 & Emission-line object (LMC) \\
LMC\,tSNE\,b2 & LEDA 148334 & Sy\,1 galaxy $z=0.0423$ & 9,10 & Emission-line galaxy $z=0.0435$ \\
LMC\,tSNE\,b3 & 2MASS J05543583$-$7101543 & QSO $z=1.88$ (photo) & 11,12 & AGN $z=1.95$ \\
LMC\,tSNE\,b4 & AT20G J050820$-$643319 & Blazar $z=0.5223$ & 13 & Blazar $z=0.522$ \\
LMC\,tSNE\,b5 & WISE J051803.01$-$632559.2 & Blazar-like radio source & 14 & AGN $z=1.50$ \\
LMC\,tSNE\,b6 & Gaia DR3 4653009410644129536 & (Thick disk star) & & Early-M dwarf \\
LMC\,tSNE\,b7 & OGLE GALAXY-LMC570.17.14 & Galaxy & 15 & ? \\
LMC\,tSNE\,b8 & 2MASS J06224188$-$6859218 & & & AGN $z=2.80$ \\
LMC\,tSNE\,b9 & PMN J0443$-$6651 & Blazar $z=0.746$ & 16,17,18 & ? \\
SMC\,PRF\,a1 & Gaia DR3 4637825429910091392 & AGN $z=1.722$ & 2 & AGN $z=1.71$ \\
SMC\,PRF\,a2 & 2MASS J00383236$-$7048470 & & & AGN $z=1.08$ \\
SMC\,PRF\,a3 & Gaia DR3 4688388744150331264 & AGN $z=1.751$ & 2 & AGN $z=1.75$ \\
SMC\,PRF\,a4 & Gaia DR3 4684665999641916928 & AGN $z=2.311$ & 2 & AGN $z=2.30$ \\
SMC\,PRF\,a5 & 2MASS J00270098$-$7357427 & & & AGN $z=2.34$ \\
SMC\,PRF\,a6 & PMN J0103$-$7006 & Radio source & 3 & AGN $z=1.65$ \\
SMC\,PRF\,g1 & 2MASS J00554909$-$7430584 & & & Galaxy $z=0.037$ \\
SMC\,PRF\,g2 & LEDA 264791 & Galaxy & 9 & Galaxy $z=0.065$ \\
SMC\,PRF\,g3 & LEDA 174103 & Galaxy & 9 & Galaxy $z=0.065$ \\
SMC\,PRF\,g4 & LEDA 265827 & Galaxy & 9 & Galaxy $z=0.089$ \\
SMC\,PRF\,g5 & 2MASX J00511677$-$7340016 & Radio galaxy & 1 & Galaxy $z=0.065$ \\
SMC\,PRF\,g6 & 2MASS J00574560$-$7427118 & & & Galaxy $z=0.080$ \\
\hline
\end{tabular}
\label{table:SALTclass}\\
\parbox{\textwidth}{Ref.:
$^1$~\citet{2002MNRAS.335.1085F};
$^2$~\citet{2023A&A...674A..41G};
$^3$~\citet{1994ApJS...91..111W};
$^4$~\citet{2011AJ....142..103B};
$^5$~\citet{2020MNRAS.495...98D};
$^6$~\citet{2013A&A...558A.101S};
$^7$~\citet{2006A&A...456..623E};
$^8$~\citet{1901ApJ....14..144P};
$^9$~\citet{2003A&A...412...45P};
$^{10}$~\citet{Jones2009};
$^{11}$~\citet{2009Gruendl};
$^{12}$~\citet{2012ApJ...747..107K};
$^{13}$~\citet{2011MNRAS.417.2651M};
$^{14}$~\citet{2014ApJS..215...14D};
$^{15}$~\citet{2012AcA....62..219S};
$^{16}$~\citet{1972AuJPh..25..581M};
$^{17}$~\citet{2019ApJS..242....4D};
$^{18}$~\citet{2021AJ....162..177P}}
\end{table*}

\begin{figure}
\centering
\begin{tabular}{c}
	\includegraphics[width=0.455\textwidth]{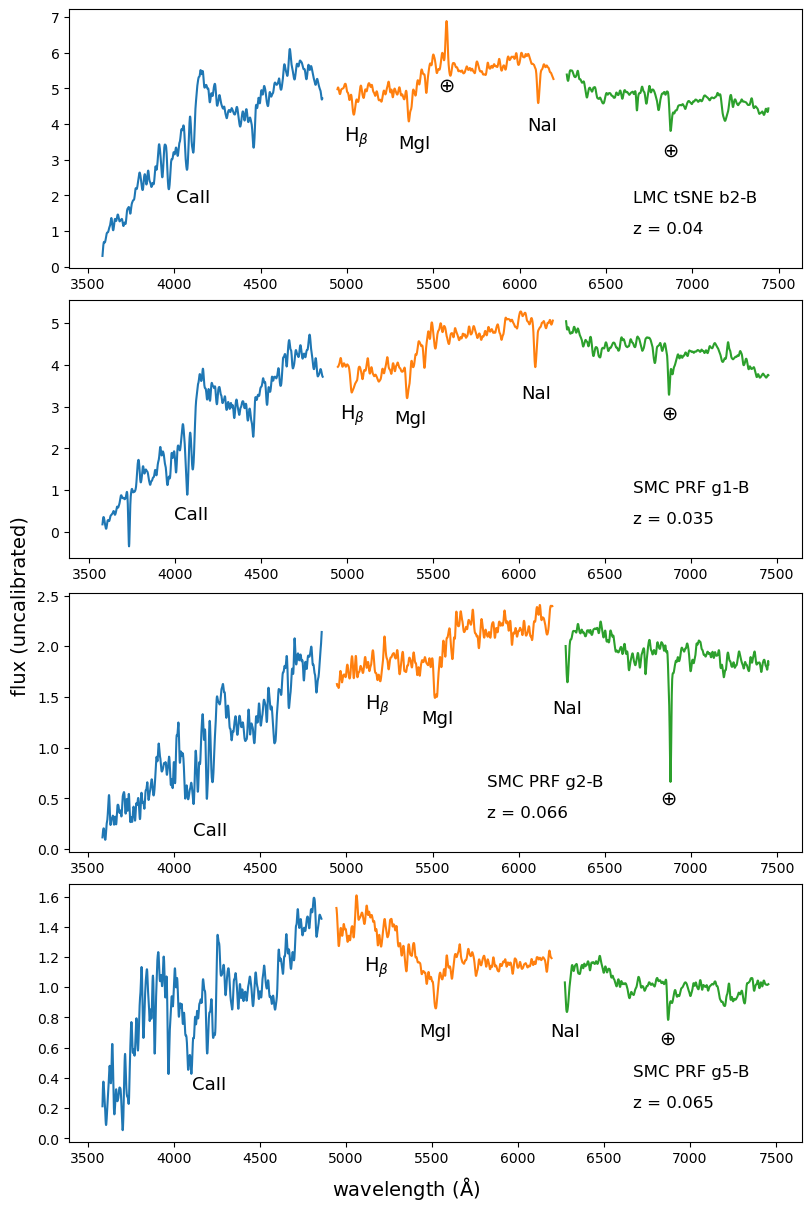}
    \end{tabular}
    \caption{Serendipitous spectra of galaxies.}
    \label{fig:serendip-gal}
\end{figure}

\begin{figure}
\centering
\begin{tabular}{c}
	\includegraphics[width=0.455\textwidth]{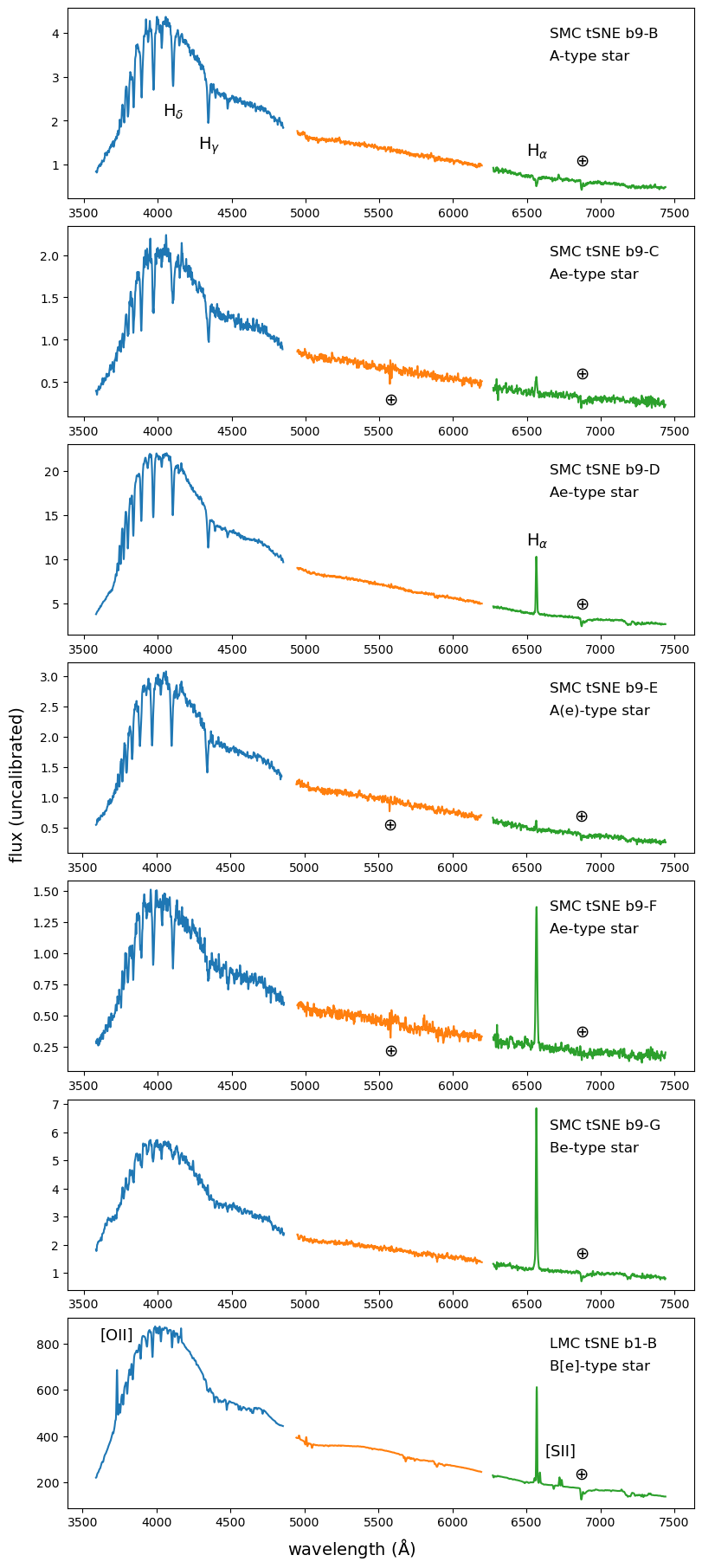}
    \end{tabular}
    \caption{Serendipitous spectra of hot (AB-type) stars most of which display line emission.}
    \label{fig:serendip-hot}
\end{figure}

\begin{figure}
\centering
\begin{tabular}{c}
	\includegraphics[width=0.455\textwidth]{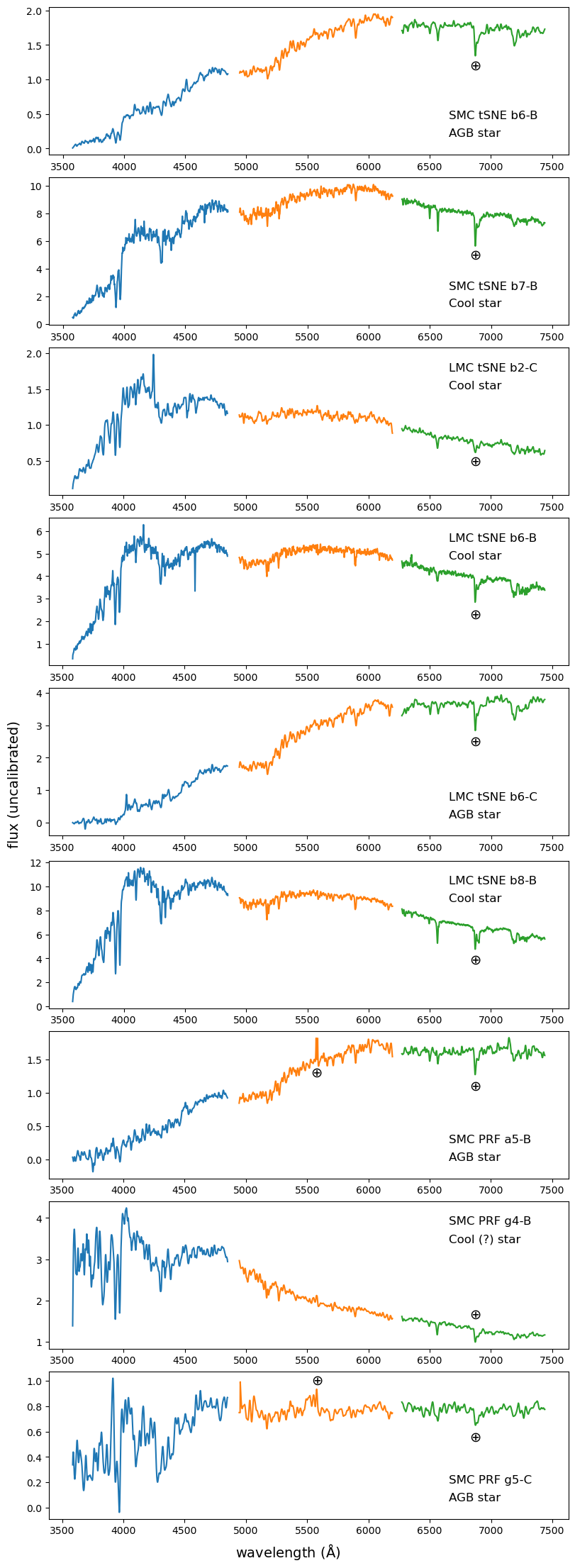}
    \end{tabular}
    \caption{Serendipitous spectra of cool stars.}
    \label{fig:serendip-cool}
\end{figure}

We also extracted spectra of additional objects on the slit, sometimes intentionally (by choosing an appropriate position angle) but at other times entirely serendipitously. The former mostly picked suspected galaxies, which turned out to be exactly that (and always at the exact same redshift as the principal target), whilst the latter invariably turned out to be cool stars, reflecting the dominant contaminant of radio selected catalogues -- except in N\,11 (near LMC\,tSNE\,b1; see below) and within 50--100 pc from the famous Be-star cluster NGC\,330 (near SMC\,tSNE\,b9) where we find A-type stars among which some display Balmer line emission. The results are summarised in table~\ref{table:serendipitous} and the spectra are displayed in figures~\ref{fig:serendip-gal}, \ref{fig:serendip-hot} and \ref{fig:serendip-cool}.

\begin{table*}
\caption{Serendipitous spectral classifications.}
\vspace{5pt}
\begin{tabular}{lllll}
\hline\hline
Target & Known source & \multicolumn{3}{c}{\rule{35mm}{0.2mm}\ Classification\ \rule{35mm}{0.2mm}} \\
 & & Literature & ref. & This work \\
 \hline
\multicolumn{5}{l}{Galaxies} \\
LMC\,tSNE\,b2-B & LEDA 3706934 & Galaxy $z=0.0433$ & 1 & Galaxy $z=0.04$ \\
SMC\,PRF\,g1-B & LEDA 3605939 & Galaxy $z=0.0352$ & 1 & Galaxy $z=0.035$ \\
SMC\,PRF\,g2-B & LEDA 265133 & Galaxy $z=0.0663$ & 2 & Galaxy $z=0.066$ \\
SMC\,PRF\,g5-B & 2MASX J00510725$-$7339218 & Galaxy & 3 & Galaxy $z=0.065$ \\
\hline
\multicolumn{5}{l}{Hot stars / emission-line objects} \\
SMC\,tSNE\,b9-B & Gaia DR3 4688992921584824064 & & & A-type star \\
SMC\,tSNE\,b9-C & Gaia DR3 4688992891430408320 & & & Ae-type star \\
SMC\,tSNE\,b9-D & [MA93] 830 & Emission-line star & 4 & Ae-type star \\
SMC\,tSNE\,b9-E & Gaia DR3 4688993093383545472 & & & A(e)-type star \\
SMC\,tSNE\,b9-F & Gaia DR3 4688992887148816000 & & & Ae-type star \\
SMC\,tSNE\,b9-G & [MA93] 903 & Be star & 5 & Be-type star \\
LMC\,tSNE\,b1-B & HD 268726 & Blue supergiant + line emission & 6 & B[e]-type star \\
%LMC\,tSNE\,b1-C & ( 4$^h$57$^m$22$\rlap{.}^s$6 $-$66$^\circ$23$^\prime$18$^{\prime\prime}$ ) & & & Ultrafast nitrogen / helium / sulfur knot \\
\hline
\multicolumn{5}{l}{Cool stars} \\
SMC\,tSNE\,b6-B & OGLE SMC-LPV-9159 & Long-period variable star & 7 & AGB star \\
SMC\,tSNE\,b7-B & 2MASS J01104014$-$7228304 & Galactic G5 star & 8 & Cool star \\
LMC\,tSNE\,b2-C & Gaia DR3 5476674461699118080 & (Halo star) & & Cool star \\
LMC\,tSNE\,b6-B & Gaia DR3 4653009174405073664 & (Halo star) & & Cool star \\
LMC\,tSNE\,b6-C & Gaia DR3 4653009827240090240 & & & AGB star \\
LMC\,tSNE\,b8-B & Gaia DR3 5279617861771350784 & (Thick disk star) & & Cool star \\
SMC\,PRF\,a5-B & OGLE SMC-LPV-470 & Long-period variable star & 7 & AGB star \\
SMC\,PRF\,g4-B & Gaia DR3 4686956011781406208 & (Thick disk star) & & Cool but blue continuum \\
SMC\,PRF\,g5-C & OGLE SMC-LPV-8747 & Long-period variable star & 7 & AGB star \\
\hline
\end{tabular}
\label{table:serendipitous}\\
\parbox{\textwidth}{Ref.:
$^1$~\citet{2011MNRAS.416.2840L};
$^2$~\citet{2012ApJS..199...26H};
$^3$~\citet{2013XMM};
$^4$~\citet{1993A&AS..102..451M};
$^5$~\citet{2007A&A...472..577M};
$^6$~\citet{1991PASP..103.1123F};
$^7$~\citet{2011AcA....61..217S};
$^8$~\citet{2010ApJ...719.1784N}}
\end{table*}

%%%%%%%%%%%%%%%%%%%%%%%%%%%%%%%%%%%%%%%%%%%%%%%%%%
\section{T-SNE map of the blazar sample}\label{blazmap}
\begin{figure*}
\centering
\begin{tabular}{c}
	\includegraphics[width=\textwidth]{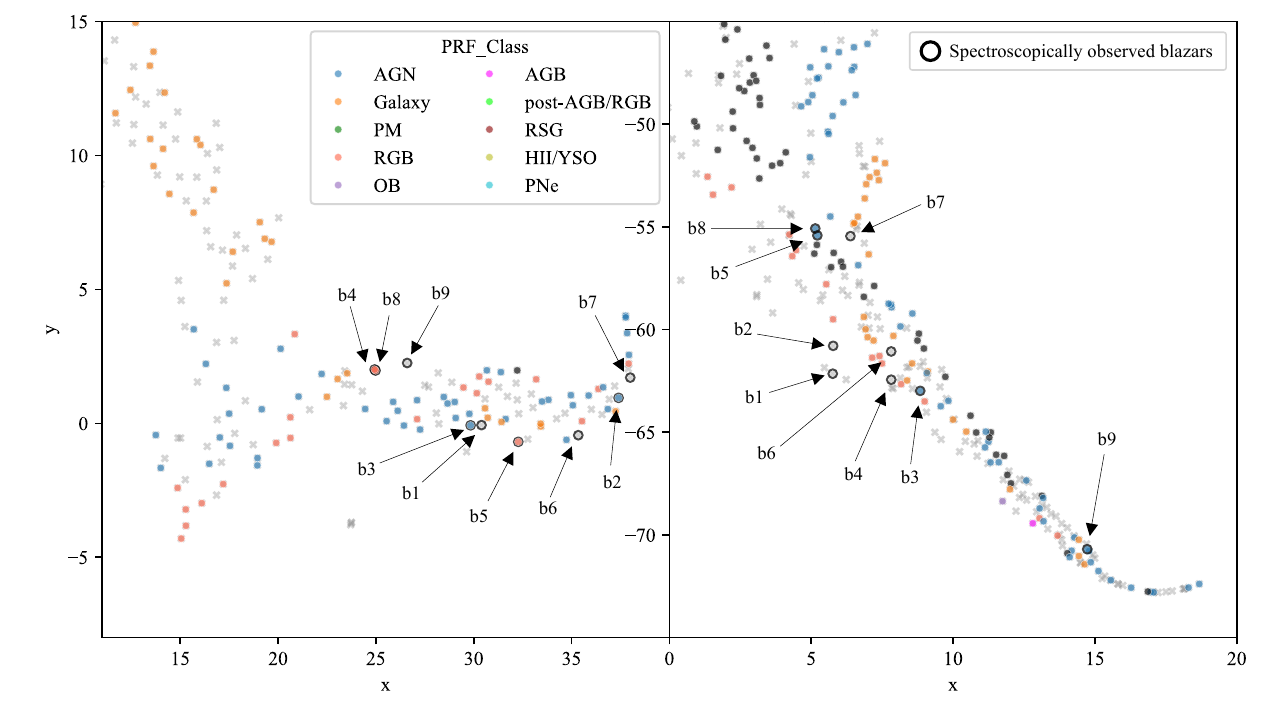}
    \end{tabular}
    \caption{Zoomed in version Figure \ref{fig:tsnePRF}, focusing on the blazar candidate sequence (as highlighted in Figure \ref{fig:MCtsneclasses}). The unfilled black circles show the locations of the spectroscopically observed blazar candidates, as outlined in Appendix \ref{SALTspec}. Classifications of the t-SNE map sources of the SMC (left) and LMC (right) are made by a PRF classifier. Grey crosses represent sources without a high probability ($>$ 80\%) for a particular class.}
    \label{fig:tsnePRF_blaz}
\end{figure*}

Figure \ref{fig:tsnePRF_blaz} shows the locations of the blazar candidates on the t-SNE map of the SMC (left) and LMC (right) that were spectroscopically observed with SALT, as outlined in Appendix \ref{SALTspec}. Also plotted are the classifications made with the PRF classifier. 

\section{VISTA variability of the blazar sample}\label{blazvar}

\begin{figure*}
\centering
\begin{tabular}{c}
	\includegraphics[width=0.8\textwidth]{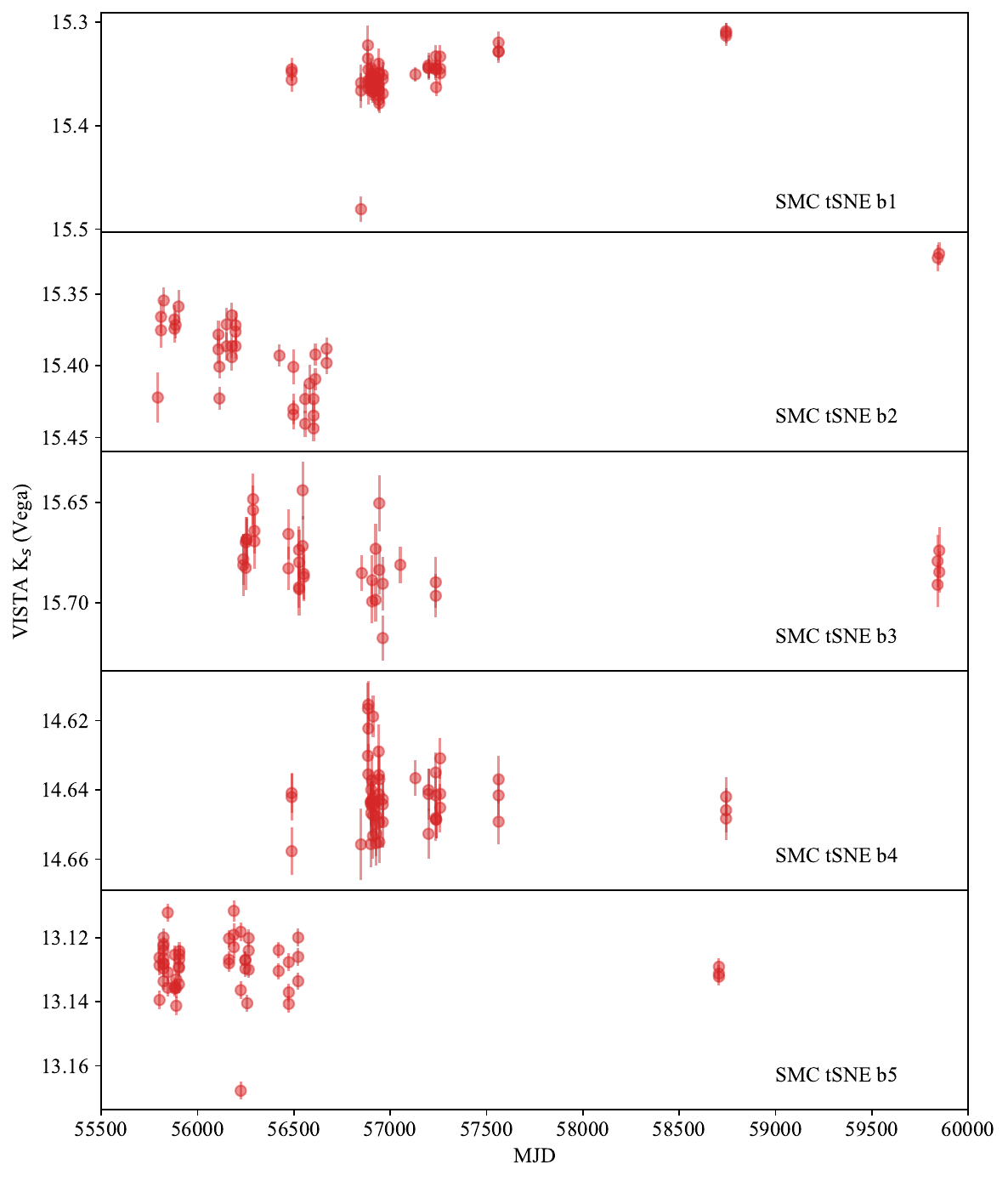}
    \end{tabular}
    \caption{The near-IR light curves of the SMC blazar sources that were not confirmed as stars in the K$_s$ band (Vega) taken from VMC DR7. }
    \label{SMCvar}
\end{figure*}

\begin{figure*}
\centering
\begin{tabular}{c}
	\includegraphics[width=0.8\textwidth]{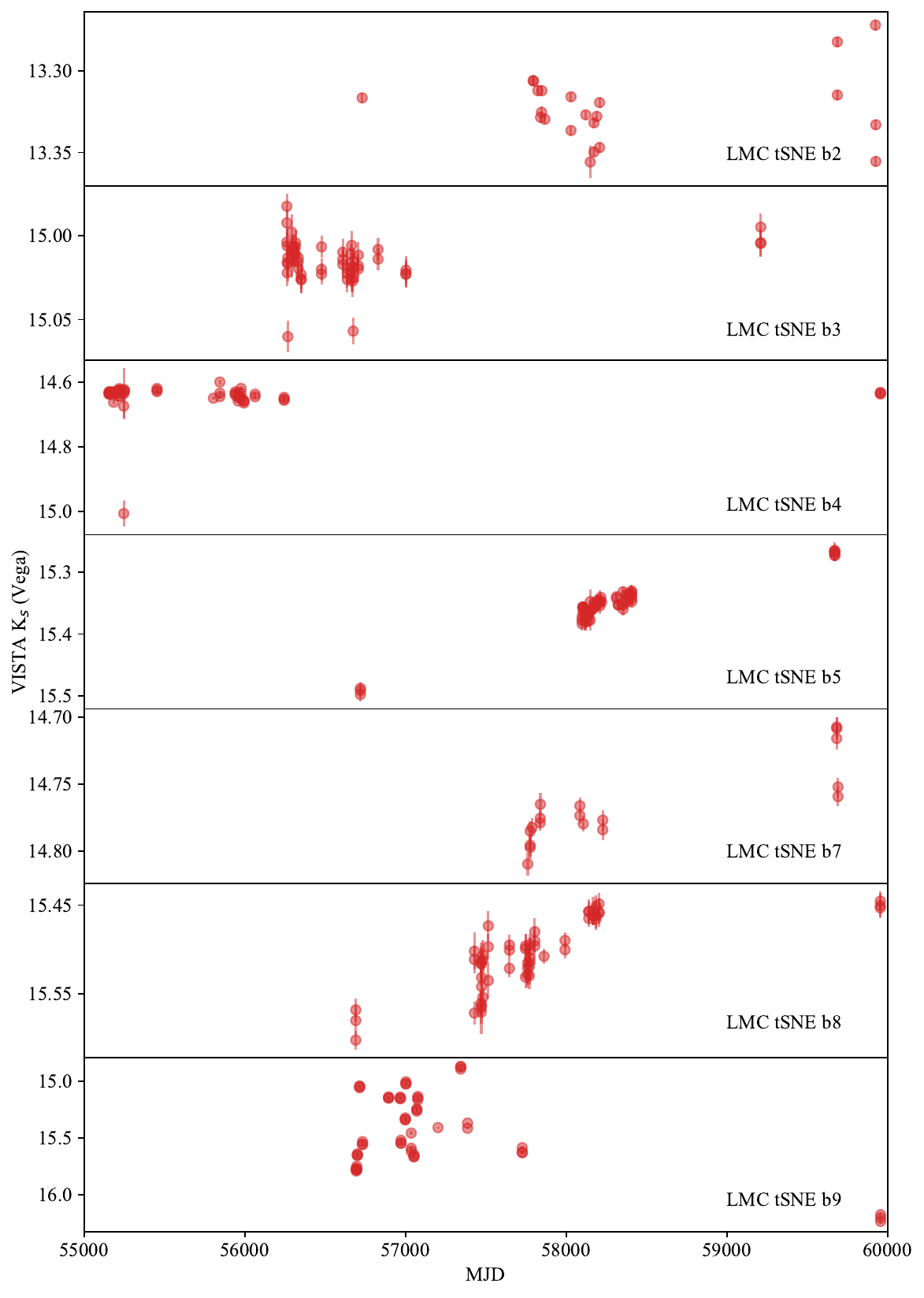}
    \end{tabular}
    \caption{The near-IR light curves of the LMC blazar sources that were not confirmed as stars in the K$_s$ band (Vega) taken from VMC DR7. }
    \label{LMCvar}
\end{figure*}

Figures \ref{SMCvar} and \ref{LMCvar} show the light curves of the 12 sources from the spectroscopically observed blazar sample (see Section \ref{blazseq} and Appendix \ref{SALTspec}) that were confirmed, via the optical spectroscopy as extragalactic or with no obvious classification due to poor signal-to-noise. The light curves were created using VISTA $K_{\rm s}$ band (Vega) from VMC DR7 \citep{2025VMCDR7}, using the Aperture 3 magnitudes, from 2009 to 2023.

Of these sources only LMC\,tSNE\,b5 and LMC\,tSNE\,b8 show the slow variation in brightness expected of an AGN, both getting brighter over time. LMC\,tSNE\,b9 shows an unexpected rapid variability over a range of $\sim$ one magnitude.
%%%%%%%%%%%%%%%%%%%%%%%%%%%%%%%%%%%%%%%%%%%%%%%%%%%

% Don't change these lines
\bsp	% typesetting comment
\label{lastpage}
\end{document}